\documentclass[twocolumn]{aastex631}
\usepackage{natbib}
\usepackage{amsmath}
\usepackage{spverbatim}

\shorttitle{VLM-Based Galaxy Merger Classification}
\shortauthors{Chiaberge et al.}

\begin{document}

\title{Vision-Language Model Ensembles Achieve Human-Expert Accuracy for Galaxy Merger Classification}


\author[0000-0003-1564-3802]{Marco Chiaberge}
\affiliation{Space Telescope Science Institute for the European Space Agency (ESA), ESA Office, 3700 San Martin Drive, Baltimore, MD, USA }
\affiliation{The William H. Miller III Department of Physics \& Astronomy, Johns Hopkins University, Baltimore, MD, USA }

\author[0000-0002-6689-505X]{Elias Stengel-Eskin}
\affiliation{The University of Texas at Austin, Department of Computer Science, Austin, Texas, USA}

\author[0000-0000-0000-0000]{Massimo Stiavelli}
\affiliation{Space Telescope Science Institute, 3700 San Martin Drive Baltimore, MD 21218, USA}
\affiliation{The William H. Miller III Department of Physics \& Astronomy, Johns Hopkins University, Baltimore, MD, USA }


\author[0000-0002-5222-5717]{Colin Norman}
\affiliation{Space Telescope Science Institute, 3700 San Martin Drive Baltimore, MD 21218, USA}
\affiliation{The William H. Miller III Department of Physics \& Astronomy, Johns Hopkins University, Baltimore, MD, USA }
\email{}

\correspondingauthor{Marco Chiaberge}
\email{marcoc@stsci.edu}

\begin{abstract}
We present a proof-of-concept study demonstrating that an ensemble of
Vision--Language Models (VLMs) combined using
a Bayesian statistical framework can classify galaxy merger morphologies
with accuracy comparable to trained human experts. We deploy
15 VLM classifier configurations, spanning four model architectures
(Gemma-4 E2B, Gemma-4 E4B, Qwen2.5-VL, and Qwen3-VL) tested with
up to four prompt engineering strategies each. We evaluate their
performance against a truth-known sample of 41 VELA+SUNRISE mock galaxy
images from \citet{Lambrides2021a}. As a proof-of-concept, all validation is performed on these mock images; application to real observational samples will require additional calibration and observational validation. The VLM ensemble achieves 
83.3\% accuracy on confident classifications (merger
probability $p_{\rm M} \ge 0.8$ or $p_{\rm M} \le 0.2$) and 58.3\% completeness, with 5
misclassified galaxies, compared to 85\% for
both human accuracy and completeness. The ensemble recovers the population merger
fraction to within $0.66\sigma$ of the truth
($f_{\rm M} = 0.52 \pm 0.09$ vs.\ true value of 0.585). 
Bayesian weighting improves overall accuracy by 17.1 percentage points over simple majority voting, with sensitivity improving by 29.2 percentage points. The VLM ensemble produces 5 misclassified galaxies (2 false positives, 3 false negatives), comparable to the 6 misclassifications (5 false positives, 1 false negative) reported for human classifiers by L21. The apparent differences in error profiles are not statistically significant given the sample size of 41 galaxies. VLMs also produce more moderate per-galaxy merger probability distributions (27\% uncertain) than the more polarized human distributions (15\% uncertain), though this difference is also consistent with statistical fluctuation. These results establish VLMs as scalable, reproducible alternatives to human classifiers within a Bayesian probabilistic merger-fraction framework, for future large-survey applications.


\end{abstract}


\section{Introduction}
\label{sec:intro}

Significant galaxy mergers throughout cosmic time play a fundamental role in theories of galaxy evolution \citep[e.g.][]{Conselice2014}. Observational estimates of merger rates have not converged, with studies using overlapping fields and similar wavelength coverage yielding disparate results \citep{Mantha2018,Duncan2019}. Robustly identifying systems undergoing ongoing (pre-coalescence) or recently completed (post-coalescence) mergers is therefore critical for constraining galaxy evolution models, particularly at cosmic noon ($1.5 \leq z \leq 2.5$) where star formation rates peak \citep{MadauDickinson2014} and where AGN activity also reaches a maximum \citep[e.g.][]{Ueda2014,Aird2015}.

Visual identification of galaxy mergers from imaging data remains a labor-intensive task traditionally performed by human experts. Human visual classification campaigns require substantial coordinated effort \citep[e.g., Galaxy Zoo;][]{Lintott2008,Willett2013}, and individual classifiers exhibit substantial variability in their merger assessments \citet[][hereinafter L21]{Lambrides2021a}. L21 demonstrated that this variability is not merely noise: the bias introduced by human classification is a function of the intrinsic merger fraction of the sample, meaning that the standard approach of comparing science and control samples under the assumption of equal bias is internally inconsistent. They proposed a Bayesian framework that explicitly measures each classifier's class-dependent accuracies using a truth-known mock sample, and propagates those measured biases into a probabilistic model to estimate the population merger fraction and per-galaxy merger probabilities.

Automated methods offer an alternative to human classification. Non-parametric morphological statistics \citep{Abraham1994,Conselice2000,Lotz2004} can exhibit misclassification rates as high as 20--30\% at high redshifts \citep{HuertasCompany2015}. Deep-learning classifiers \citep{Dieleman2015,Ackermann2018,Pearson2019,
Ciprijanovic2020,Ciprijanovic2021} achieve lower misclassification rates
within their training domain but require task-specific training and do
not generalize reliably across redshift or survey depth. For example,
\citet{Pearson2019} found that a CNN trained on SDSS observations
achieved 91.5\% accuracy on that same data set, but accuracy dropped to
64.6\% when the network was applied across data sets, and to only 53.0\%
when classifying simulated EAGLE images with the observation-trained
network. Machine-learning classifiers trained on human-labeled data also
inherit any systematic biases present in the training set. These concerns extend to general-purpose VLMs, whose behavior
reflects biases acquired during large-scale pre-training rather
than being intrinsically ``unbiased'' classifiers. Unlike
task-specific deep-learning classifiers, which learn to recognize
mergers purely from human-labeled example images, VLMs are
pretrained on vast, mixed text-and-image corpora that likely
include scientific and educational content describing astronomical
concepts such as galaxy mergers, tidal tails, and interacting
systems. As a result, a VLM's classification of a given galaxy
image may be shaped not only by learned visual pattern recognition
but also by prior conceptual and linguistic associations with the
term ``merger'' acquired from text, independent of any direct,
per-image human labeling. 

Recent work has advanced simulation-driven and machine-learning merger
classifiers beyond these early approaches. For example,
\citet{Ferreira2024} introduce a hybrid convolutional-neural-network and
Vision Transformer ensemble trained on realism-added IllustrisTNG mock
observations to identify mergers and distinguish interacting pairs from
post-mergers in UNIONS imaging. \citet{DeGraaff2025} use adaptive
deep-learning models combined with morphological parameters to classify
merger stages in realistic simulations. The Galaxy Merger Challenge
\citep{MargalefBentabol2024} benchmarks several state-of-the-art
machine-learning methods across multiple simulations and Hyper
Suprime-Cam data, highlighting reasonably strong in-domain performance
(precision $\sim$70--80\%, recall $\sim$70--77\% for binary merger/non-merger
classification) alongside a substantial recall degradation under domain shift
($\sim$20--40\%) and poor performance on the finer-grained multi-class
(pre-/post-/non-merger) task.

Forthcoming surveys make these methodological requirements increasingly demanding in practice. Current HST legacy programs such as CANDELS \citep{Grogin2011,Koekemoer2011} already provide $\sim 50{,}000$ galaxies to $z\sim 3$; JWST programs such as CEERS \citep{Finkelstein2023} and PRIMER \citep{Donnan2024} extend morphological coverage to $z>4$; Euclid and the Vera Rubin Observatory will further increase the surveyed volume of galaxy samples, while the planned Roman Space Telescope \citep{Spergel2015} High Latitude Wide Area Survey, which will deliver HST-like near-infrared imaging over $\sim 2{,}400\ \mathrm{deg}^2$, enabling morphological studies of hundreds of millions of galaxies across cosmic time. In the radio, SKA surveys will need similar morphological classifications. At scales of millions of galaxies, assembling a stable human classifier panel with individually calibrated accuracy measurements following the L21 framework is logistically impractical.
More fundamentally, a classifier validated only against other human labels is anchored to the same subjective judgments that a simulation-based bias-correction framework is designed to overcome, making the combination of a reproducible algorithmic ensemble with a truth-known mock calibration set not just convenient but necessary.

Recent work in multimodal AI has shown that systematic disagreement among diverse visual agents carries exploitable information about classification difficulty: \citet{Sivakumaran2025} demonstrate that multi-agent disagreement can be leveraged to identify ambiguous cases and improve ensemble decisions for tool-using vision–language models. In this work, we exploit this principle in a new context: we deploy an ensemble of general-purpose, instruction-following Vision-Language Models (VLMs --- large language models with native multimodal vision capabilities) whose individual classifiers deliberately span a range of classification behaviors through prompt variation, from merger-biased to isolated-biased, rather than converging on a single classification strategy. Rather than resolving inter-classifier disagreement directly, we channel it through the Bayesian framework of L21, which models each classifier's individual bias pattern explicitly and uses it as a calibrated weight in the merger fraction likelihood. Our central question is whether this combination --- a diverse VLM ensemble embedded in a bias-aware probabilistic framework --- can serve as a viable substitute for human classifiers for galaxy merger identification, and if so, whether it can be scaled to the very large galaxy samples expected from current and future surveys (HST legacy fields, JWST, Roman Space Telescope).

The present paper is framed as a proof-of-concept calibration study. 
Because scientific reproducibility requires that identical inputs yield identical classifications
over time, we restrict ourselves to locally runnable, open-weight VLMs with fixed checkpoints,
rather than API-based frontier models whose weights and behaviors can change without notice.
We use the same class of VELA+SUNRISE truth-known mock galaxy sample \citep{Ceverino2014,Ceverino2015,Simons2019}
employed in the L21 framework to measure how accurately small, locally runnable VLMs can reproduce merger versus
non-merger classifications. We then apply the full Bayesian ensemble machinery of L21 and compare the resulting
classification accuracy to the 14 human classifiers reported in that work.
As a consequence, all validation in this study is performed on these truth-known mock galaxy images.
For datasets with matching imaging characteristics (e.g., the 3D-HST GOODS-S observations analyzed in \citealt{Lambrides2021b}),
this mock-based calibration can be applied directly without additional retraining.
For surveys with substantially different noise properties, depth, or filter coverage, recalibration on domain-matched
mock images will be required.

This paper is organized as follows. Section~\ref{sec:data} describes the mock galaxy sample. Section~\ref{sec:methods} presents the VLM ensemble and Bayesian methods. Section~\ref{sec:results} reports the results. Section~\ref{sec:discussion} discusses implications and limitations. Section~\ref{sec:conclusions} summarizes our conclusions.
Throughout, the research was
conducted using a human-directed AI collaboration model, with large language models assisting in
code development, statistical analysis, and drafting under close human supervision; this workflow is
described in more detail in Appendix~\ref{sec:appendix_llm_use}.

\section{Data: Mock Galaxy Sample}
\label{sec:data}

We validate our VLM classifier ensemble using the same mock galaxy images used in L21. More details can be found in the original paper by those authors. The images are taken from the VELA zoom-in hydrodynamical simulations \citep{Ceverino2015} processed through the SUNRISE dust-radiative transfer code \citep{Snyder2015,Simons2019}. 
As in L21, we treat both pre-coalesced and post-coalesced systems as mergers.
In that framework, a mock galaxy is labeled as merger if either it underwent a significant merger within the
previous 100 Myr, or it has a companion within 35 kpc. Mock Hubble ACS/WFC3 images were created for each galaxy in three bands (ACS/WFC F435W, ACS/WFC F775W, WFC3-IR F160W) and noise-added to simulate the 3DHST reduction of the GOODS-S field (Simons et al. 2019). 
Each galaxy is presented to the VLM classifiers as a two-panel composite image (Figure 4 and
Appendix~\ref{sec:appendix_example_merger}) showing the WFC3/IR F160W near-infrared band and an RGB composite
(r: F160W, g: F775W, b: F435W); Appendix~\ref{sec:appendix_example_merger} provides a worked example of the
resulting VLM classification output for one such system.

Our working sample comprises 41 galaxies with known intrinsic merger states: 24 true mergers and 17 isolated systems. Following L21, we additionally include a set of 9 fake mergers (i.e. close projected pairs created by superimposing images of two mock isolated galaxies with a random separation of less than 8$\arcsec$) to assess how well classifiers distinguish chance alignments from true interactions. The fake mergers are included in the computation of individual classifier accuracies (merger recall r$_M$, non-merger recall r$_i$) but are excluded from the merger fraction analysis, to exactly replicate the same experiment as in the original work utilizing human classifiers.

The true merger fraction of our working sample is f$_M$ = 24/41 = 0.585. The galaxy IDs, redshifts (spanning z = 1.0--3.5), and intrinsic types are drawn from the same mock galaxy catalog used in L21. The use of the same simulation suite enables a direct, controlled comparison between VLM and human classifier performance.

\section{Methods}
\label{sec:methods}

\subsection{VLM Classifier Ensemble}
\label{sec:vlm_ensemble}

We employ an ensemble of 15 VLM classifier configurations to assess galaxy merger status. The ensemble comprises four VLM architectures (see Appendix~\ref{sec:appendix_impl} for more details) --- Gemma-4 E2B (2B effective parameters), Gemma-4 E4B (4B effective parameters), Qwen2.5-VL 7B, and Qwen3-VL 4B --- evaluated with up to four prompt engineering strategies each: balanced, open, optimal, and strict. Qwen2.5-VL uses only three strategies (open, optimal, strict), yielding 15 total configurations (Table 1). 
All models use 4-bit quantization (Q4\_K\_M or Q4\_K\_XL) for efficient local inference via
\texttt{llama.cpp}; additional implementation and inference details, including model files and
runtime configuration, are provided in Appendix~\ref{sec:appendix_impl}.
A key design choice in this proof-of-concept is to work at effectively zero marginal cost using only small, publicly available models that can run on a single consumer-grade machine. 
All inference is performed locally, without commercial APIs or cloud GPU resources, so that the
entire pipeline can be reproduced on standard consumer hardware; using fixed, locally hosted
checkpoints avoids the version drift inherent to API-based models.
This deliberately restricts us to modest VLM architectures, and the resulting performance should therefore be interpreted as a conservative baseline for what can be achieved with off-the-shelf models under minimal computational assumptions. This ensemble size of 15 classifiers is directly comparable to the 14 human classifiers employed by L21, enabling a robust apples-to-apples comparison between VLM and human classification performance.

\begin{table*}
\caption{VLM Classifier Configurations}
\label{tab:vlm_configs}
\centering
\begin{tabular}{ccccc}
\hline
ID & Model       & Eff.\ Parameters & Quantization           & Prompt Strategies \\
\hline
1--4   & Gemma-4 E2B & 2B & 4-bit (Q4\_K\_XL) & balanced, open, optimal, strict \\
5--8   & Gemma-4 E4B & 4B & 4-bit (Q4\_K\_M)  & balanced, open, optimal, strict \\
9--11  & Qwen2.5-VL  & 7B & 4-bit (Q4\_K\_M)  & open, optimal, strict \\
12--15 & Qwen3-VL    & 4B & 4-bit (Q4\_K\_M)  & balanced, open, optimal, strict \\
\hline
\end{tabular}
\tablecomments{E2B and E4B refer to the 2B and 4B effective-parameter Gemma-4 on-device models, respectively. Further implementation details are provided in Appendix~\ref{sec:appendix_impl}}
\end{table*}

Each VLM classifier independently evaluates galaxy images and assigns a binary classification: merging or non-merging. The merging class encompasses major mergers, minor mergers, and major disturbances, while the non-merging class includes minor disturbances and systems showing no evidence of gravitational interactions (Table 2). This classification scheme follows the framework of L21 for direct comparability.

\begin{table*}
\caption{Five-Category Primary Classification Schema and Binary Collapse}
\label{tab:class_schema}
\centering
\begin{tabular}{lll}
\hline
Primary Label        & Collapsed Label & Description \\
\hline
no\_evidence         & non-merging     & Smooth, symmetric morphology; single nucleus; no tidal features \\
disturbance\_minor   & non-merging     & Single core; mild, ambiguous asymmetry or clumpiness; no clear tidal features \\
disturbance\_major   & merging         & Single core; strong asymmetry, tidal tail, or irregular envelope \\
merging\_minor       & merging         & Two resolved cores; clearly unequal mass ratio \\
merging\_major       & merging         & Two comparable bright nuclei in close proximity or shared envelope \\
\hline
\end{tabular}
\tablecomments{Primary labels follow the scheme of L21, with a binary collapse into merging vs.\ non-merging for the purposes of this analysis.}
\end{table*}

\subsection{Prompt Engineering Strategy}
\label{sec:prompts}

Prompt engineering is the primary lever for controlling model behavior. All prompts require the model to output a structured JSON object with fixed fields (primary\_label, collapsed\_label, num\_apparent\_galaxies, confidence, notes\_evidence, reasoning\_summary, and warnings). Structured output is essential for automated parsing and prevents models from hedging with free-form prose. We emphasize that this strategy is an essential aspect of this project.

A central design goal is to force each model to base its classification strictly on observable morphological evidence rather than prior probability or stylistic preference. The prompt explicitly instructs the model to enumerate specific visual features (tidal tails, multiple nuclei, asymmetry, chaotic star formation) as positive evidence before assigning a label. Models without this instruction default to no\_evidence for ambiguous cases, inflating false negatives.

We evaluate four prompt variants designed to elicit systematically different classification behaviors. The optimal prompt was developed first through iterative refinement, encoding specific morphological criteria intended to achieve the best single-model performance. From this baseline, we created the open and strict variants to deliberately span a range of classification behaviors: the open prompt requires only one suggestive morphological feature to trigger a merging classification, favoring merger detection at the expense of false positives, while the strict prompt requires two or more unambiguous features, favoring reliable detections at the expense of missed mergers. After initial testing revealed that even the optimal prompt produced merger-biased behavior for some models, we introduced the balanced prompt as a deliberately neutral variant: it symmetrically emphasizes both merger and non-merger evidence, removes language that could bias the model toward either class, and explicitly instructs the model to enumerate features supporting both classifications before deciding. This four-strategy framework ensures that the ensemble spans a wider and more regular range of classification behaviors than any single prompt could achieve. As discussed in Section \ref{sec:disc_5_3}, the actual classification behavior depends on the interaction between prompt and model architecture, and the intended effect is not always achieved.

Two independent prompt families were developed: one anchored to the Gemma-4 E2B model and one for the Qwen models. The same four-strategy framework was applied to both Gemma families and Qwen3. For Qwen2.5-VL we also tested a balanced prompt variant, but its performance was nearly indistinguishable from the optimal prompt, so we retain only the open, optimal, and strict strategies for the final ensemble.

\subsection{Bayesian Statistical Framework}
\label{sec:bayes}

We adopt the Bayesian statistical framework of L21 to combine classifications from multiple VLM classifiers while accounting for their individual accuracies. For each classifier $i$, we define two accuracy parameters: the sensitivity $r_{M,i}$ (probability of correctly identifying a true merger) and the specificity $r_{I,i}$ (probability of correctly identifying a true isolated galaxy).

The classifier accuracies are estimated directly from performance on the full mock galaxy sample including fake mergers, following L21 (their Equations 13 and 14). For a classifier with $n_M$ true mergers correctly identified out of $N_M$ total true mergers, and $n_I$ true isolated galaxies correctly identified out of $N_I$ total isolated systems (including fakes):

\begin{equation}
r_{{\rm M},i} = \frac{n_{{\rm M},i}}{N_{{\rm M},i}} ,
\label{eq:rM}
\end{equation}

\begin{equation}
r_{{\rm I},i} = \frac{n_{{\rm I},i}}{N_{{\rm I},i}} .
\label{eq:rI}
\end{equation}

These fixed accuracy values are then used in a Bayesian model to estimate
the population merger fraction $f_{\rm M}$ and per-galaxy merger probabilities
$p_{{\rm M},j}$. The likelihood for the observed classifications,
marginalizing over the unknown true state of each galaxy
(Lambrides et al. 2021a, cf. Equations 17--18), is

\begin{multline}
P(\mathrm{obs} \mid f_{\rm M}, \{r_{{\rm M},i}\}, \{r_{{\rm I},i}\}) = \prod_j \bigg[
f_{\rm M} \prod_i P(\mathrm{obs}_{ij} \mid \mathrm{merger})
\\
+ (1 - f_{\rm M}) \prod_i P(\mathrm{obs}_{ij} \mid \mathrm{isolated})
\bigg] .
\end{multline}

Here
$P(\mathrm{obs}_{ij} = \mathrm{merger} \mid \mathrm{true\ merger})
= r_{{\rm M},i}$ and
$P(\mathrm{obs}_{ij} = \mathrm{merger} \mid \mathrm{true\ isolated})
= 1 - r_{{\rm I},i}$. 
We implement the Bayesian model in Stan, interfaced via R \citep{RCoreTeam2024,Stan2024}, and run four Hamiltonian Monte Carlo chains with 2000 iterations per chain and 1000 warmup iterations. Convergence was verified using the
Gelman--Rubin diagnostic ($\hat{R} = 1.002$ for $f_{\rm M}$) and effective
sample size ($n_{\rm eff} > 1600$), both indicating excellent convergence.

The per-galaxy merger probability $p_M$ is computed via Bayes' theorem (L21, Equation 18):

\begin{multline}
p_{{\rm M},j}
= P(\mathrm{merger} \mid \mathrm{obs}_j)
\\
= \frac{
    f_{\rm M} \prod_i P(\mathrm{obs}_{ij} \mid \mathrm{merger})
}{
    f_{\rm M} \prod_i P(\mathrm{obs}_{ij} \mid \mathrm{merger})
    + (1 - f_{\rm M}) \prod_i P(\mathrm{obs}_{ij} \mid \mathrm{isolated})
} .
\end{multline}

Following L21, we define confident classifications as those with $p_M \geq 0.8$ (confident mergers) or $p_M \leq 0.2$ (confident isolated); intermediate values ($0.2 < p_M < 0.8$) are considered uncertain. The accuracy is defined as the fraction of confident classifications that are correct.

\subsection{Treatment of Classifier Accuracies}
\label{sec:accuracies}

The Bayesian framework requires specification of classifier accuracies $r_{M,i}$ (sensitivity) and $r_{I,i}$ (specificity) for each classifier. We fix these accuracies to their maximum-likelihood estimates from the calibration sample (Equations~\ref{eq:rM}--\ref{eq:rI}), treating them as known constants in the MCMC inference. The only free parameter sampled is the merger fraction $f_M$.

We adopt this fixed-accuracy approach for the following reason. The mock galaxy calibration sample is approximately balanced between mergers and isolated systems (24 mergers, 26 isolated), providing unbiased estimates of classifier performance on both classes. Given these well-measured accuracies, the observed vote patterns on any sample---whether the same mock galaxies or a new sample of real galaxies---can be used to infer the merger fraction $f_M$ without ambiguity.

In this fixed-accuracy formulation, the classifier sensitivities
and specificities are properties of the classifier--data
combination rather than the sample's merger fraction, and should
therefore, in principle, remain valid when applied to a sample with
a different underlying merger fraction than the calibration sample.

To test this claim empirically, we performed a subsampling
experiment: starting from the 41-galaxy working sample, we
progressively removed true mergers (up to 20 of 24) to construct
sub-samples spanning true merger fractions from 0.585 down to 0.19,
while holding classifier accuracies ($r_{M,i}$, $r_{I,i}$) fixed at
their values from the full calibration sample. We reran the
fixed-accuracy Bayesian inference
on each sub-sample, averaging over 10 random removal draws per
sample size, and report uncertainties as the
16th--84th percentile spread of the posterior, consistent with our
main result. As shown in Table~\ref{tab:fM_robustness}, the
inferred merger fraction tracks the true value to within
$0.66\sigma$ at every sample size tested, and the posterior
uncertainty grows only modestly, from $+0.09/-0.10$ at $N=41$ to
$+0.14/-0.11$ at $N=21$, rather than degrading sharply. This
confirms empirically, within the mock dataset, that $r_{M,i}$ and
$r_{I,i}$ remain valid across a substantial range of merger
fractions differing from that of the calibration sample. Testing this robustness on mock images matched to surveys with
substantially different observing characteristics, or on real
observational datasets, is left to future work.

\begin{table}
\caption{Merger fraction recovery for subsamples with varying true merger
fraction, holding classifier accuracies fixed at calibration values.}
\label{tab:fM_robustness}
\centering
\begin{tabular}{ccccc}
\hline
$N$ & True $f_M$ & Inferred $f_M$ & Uncertainty & Deviation \\
\hline
41 & 0.585 & 0.523 & $+0.09/-0.10$ & 0.66$\sigma$ \\
37 & 0.541 & 0.488 & $+0.10/-0.10$ & 0.52$\sigma$ \\
33 & 0.485 & 0.442 & $+0.11/-0.11$ & 0.44$\sigma$ \\
29 & 0.414 & 0.424 & $+0.12/-0.11$ & 0.23$\sigma$ \\
25 & 0.320 & 0.335 & $+0.12/-0.12$ & 0.24$\sigma$ \\
21 & 0.190 & 0.228 & $+0.14/-0.11$ & 0.31$\sigma$ \\
\hline
\end{tabular}
\tablecomments{$N$: subsample size, obtained by removing true mergers from the
41-galaxy sample while retaining all 17 isolated galaxies; each row averages 10
random removal draws. Inferred $f_M$ and Uncertainty are the across-draw means
of the posterior median and of the 84th/16th-percentile offsets. Deviation is
the mean of each draw's own $|\hat{f}_M - f_{M,\mathrm{true}}|/\sigma_{f_M}$.}
\end{table}

The alternative approach, implemented by L21, samples $r_{M,i}$ and $r_{I,i}$ as parameters with informative Beta priors derived from calibration. While this ostensibly propagates calibration uncertainty into the inference, it introduces a degeneracy: the observed vote patterns cannot distinguish between a high merger fraction and high classifier sensitivity, since both produce more ``merger'' votes. By fixing accuracies to their calibration values, we anchor the inference to an external measurement and break this degeneracy.

We verify our choice by comparing results from both approaches in Section~\ref{sec:results_fixed_vs_sampled}.

%
%

\section{Results}
\label{sec:results}

\subsection{Individual VLM Classifier Accuracies}
\label{sec:results_4_1}

\begin{figure*}
    \centering
    \includegraphics[width=\linewidth]{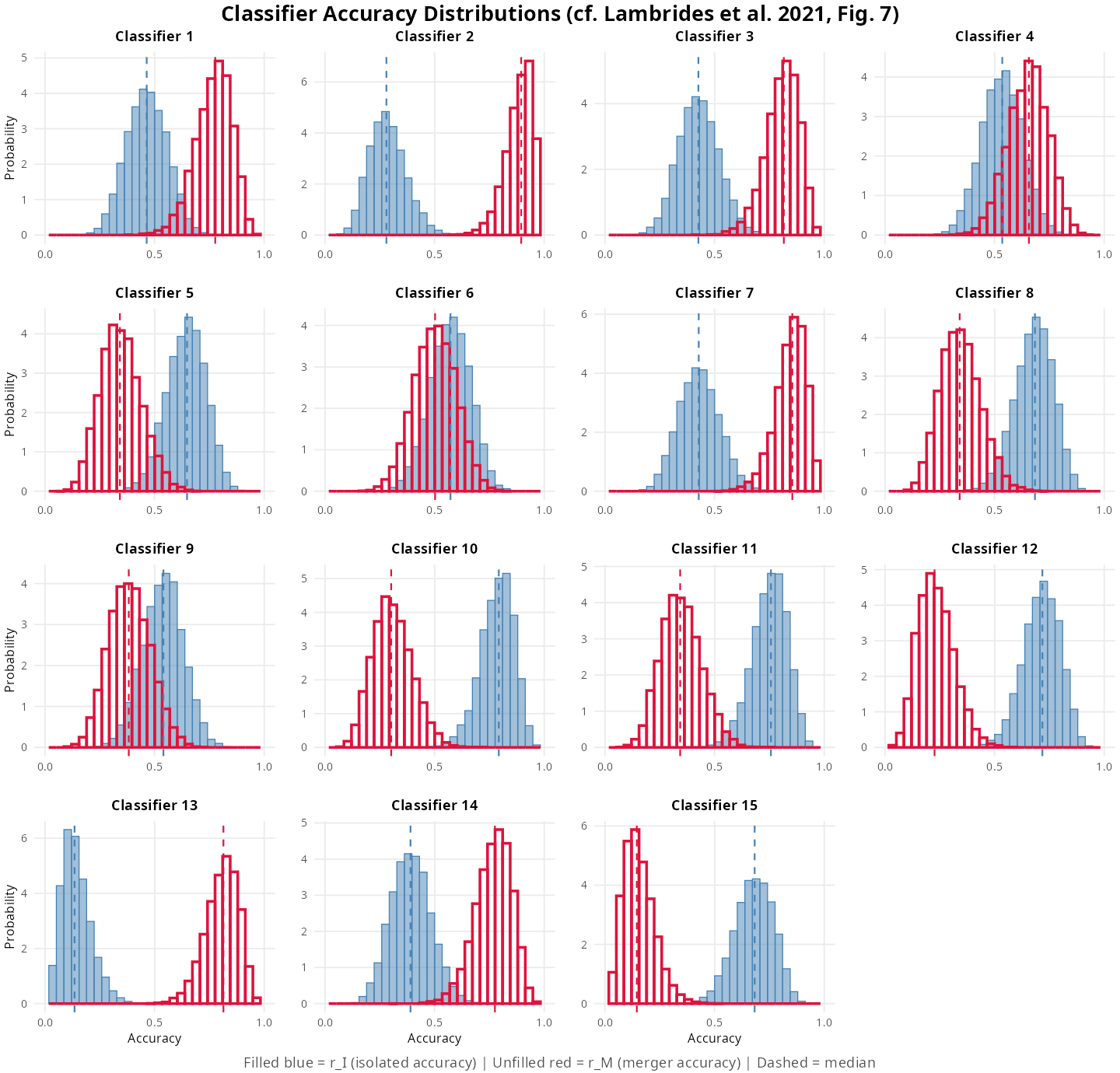}
    \caption{
    Distribution of merger recall $r_{\rm M}$ and non-merger recall $r_{\rm I}$ for the 15 VLM classifiers.
    The ensemble spans a broad range of sensitivity--specificity trade-offs, a key ingredient for Bayesian weighting.
    }
    \label{fig:rM_rI_distribution}
\end{figure*}

Table~\ref{tab:vlm_performance} presents the sensitivity ($r_M$), specificity ($r_I$), and raw accuracy for each of the 15 VLM classifier configurations. Although we refer to these four variants as ``open,'' ``balanced,'' ``optimal,'' and ``strict,'' these labels describe design intent within the two anchor models (Gemma-4 E2B and Qwen2.5-VL) rather than architecture-independent operating points, and the realized $r_M$ and $r_I$ values can shift when the same prompts are applied to Gemma-4 E4B or Qwen3-VL. The classifiers exhibit a diverse range of accuracies, with mean sensitivity $\langle r_M \rangle = 54.7\%$ and mean specificity $\langle r_I \rangle = 53.8\%$. Individual sensitivities range from 12.5\% (Qwen3-VL strict) to 91.7\% (Gemma-4 E2B open), while specificities range from 11.5\% (Qwen3-VL open) to 80.8\% (Qwen2.5-VL optimal). This diversity mirrors the variation observed among the 14 human classifiers in L21 and underscores the importance of the Bayesian weighting scheme (Figure \ref{fig:rM_rI_distribution}).

\begin{table*}
\caption{Individual VLM Classifier Performance}
\label{tab:vlm_performance}
\centering
\begin{tabular}{cccccccc}
\hline
ID & Model       & Prompt   & $r_M$  & $r_I$  & FP & FN & Acc. \\
\hline
1  & Gemma-4 E2B & balanced & 0.792 & 0.462 & 14 & 5  & 0.620 \\
2  & Gemma-4 E2B & open     & 0.917 & 0.269 & 19 & 2  & 0.580 \\
3  & Gemma-4 E2B & optimal  & 0.833 & 0.423 & 15 & 4  & 0.620 \\
4  & Gemma-4 E2B & strict   & 0.667 & 0.538 & 12 & 8  & 0.600 \\
5  & Gemma-4 E4B & balanced & 0.333 & 0.654 & 9  & 16 & 0.500 \\
6  & Gemma-4 E4B & open     & 0.500 & 0.577 & 11 & 12 & 0.540 \\
7  & Gemma-4 E4B & optimal  & 0.875 & 0.423 & 15 & 3  & 0.640 \\
8  & Gemma-4 E4B & strict   & 0.333 & 0.692 & 8  & 16 & 0.520 \\
9  & Qwen2.5-VL  & open     & 0.375 & 0.538 & 12 & 15 & 0.460 \\
10 & Qwen2.5-VL  & optimal  & 0.292 & 0.808 & 5  & 17 & 0.560 \\
11 & Qwen2.5-VL  & strict   & 0.333 & 0.769 & 6  & 16 & 0.560 \\
12 & Qwen3-VL    & balanced & 0.208 & 0.731 & 7  & 19 & 0.480 \\
13 & Qwen3-VL    & open     & 0.833 & 0.115 & 23 & 4  & 0.460 \\
14 & Qwen3-VL    & optimal  & 0.792 & 0.385 & 16 & 5  & 0.580 \\
15 & Qwen3-VL    & strict   & 0.125 & 0.692 & 8  & 21 & 0.420 \\
\hline
-- & --          & Mean     & 0.547 & 0.538 & -- & -- & 0.543 \\
\hline
\end{tabular}
\tablecomments{$r_M$ = merger recall (sensitivity); $r_I$ = non-merger recall (specificity); FP = false positives (isolated galaxies classified as mergers); FN = false negatives (mergers classified as isolated); Acc.\ = overall accuracy, $(N - \mathrm{FP} - \mathrm{FN})/N$. All quantities are computed on the 50-galaxy calibration sample (24 mergers, 26 non-mergers including 9 fake mergers), identical to the sample used by L21. Individual sensitivities range from 12.5\% (Qwen3-VL strict) to 91.7\% (Gemma-4 E2B open), while specificities range from 11.5\% (Qwen3-VL open) to 80.8\% (Qwen2.5-VL optimal).}
\end{table*}

The specific prompt variant strongly modulates the sensitivity and specificity of each VLM classifier. Open prompts tend to produce higher sensitivity but lower specificity (merger-biased), while strict prompts yield higher specificity but lower sensitivity (isolated-biased). The Gemma-4 family responds predictably and approximately monotonically to prompt strictness, making it the most controllable family for calibration purposes. Qwen3-VL is highly instruction-following but exhibits extreme swings across prompt variants (an $r_M$–$r_I$ dynamic range of about 71 percentage points), whereas Qwen2.5-VL shows minimal sensitivity to prompt framing, with its behavior dominated by internal priors rather than the detailed wording of the instructions.

\subsection{Merger Fraction Recovery}
\label{sec:results_4_2}

\begin{figure*}
    \centering
    \includegraphics[width=0.5\linewidth]{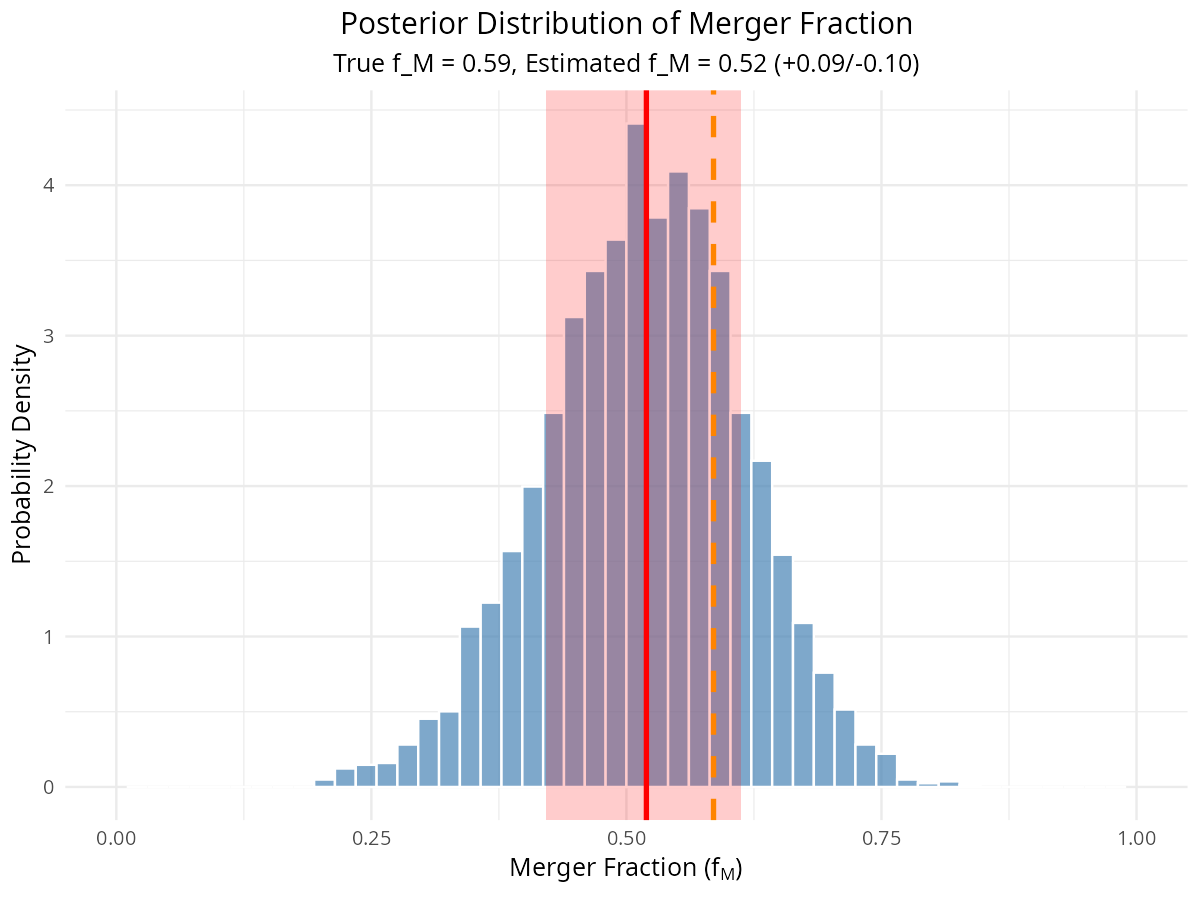}
    \caption{
    Posterior distribution of the merger fraction $f_{\rm M}$ inferred from the VLM ensemble.
    The true value (vertical line) lies well within the 68\% credible interval.
    }
    \label{fig:posterior_fM}
\end{figure*}

The Bayesian ensemble successfully recovers the true merger fraction of the mock galaxy sample. We estimate $f_{\rm M} = 0.52~(+0.09 / -0.10)$, which is within $0.66\sigma$ of the true merger fraction of 0.585 (24/41 galaxies). Figure \ref{fig:posterior_fM} shows the posterior distribution of the merger fraction, demonstrating that the true value falls well within the 68\% credible interval (0.43–0.62). This result is directly comparable to the recovery by L21, who obtained $f_{\rm M} = 0.52 \pm 0.08$ for the same true fraction of 0.585 using 14 human classifiers on the same 41 galaxies.

\subsection{Classification Performance: Majority Vote vs. Bayesian Weighting}
\label{sec:results_4_3}

Table~\ref{tab:mv_vs_bayes} compares the classification performance before (majority vote) and after (Bayesian weighting) applying the statistical framework to the 41-galaxy working sample.

\begin{table*}
\caption{Ensemble Classification Performance: Majority Vote vs. Bayesian Weighting}
\label{tab:mv_vs_bayes}
\centering
\begin{tabular}{cccc}
\hline
Metric & Majority Vote & Bayesian Weighted & Change \\
\hline
Overall Accuracy    & 65.9\% (27/41) & 82.9\% (34/41) & +17.1 pp \\
Sensitivity ($r_M$) & 58.3\% (14/24) & 87.5\% (21/24) & +29.2 pp \\
Specificity ($r_I$) & 76.5\% (13/17) & 76.5\% (13/17) & +0.0 pp \\
\hline
\end{tabular}
\tablecomments{Performance metrics computed on the 41-galaxy sample (24 mergers, 17 isolated). Bayesian classification uses the threshold $p_M > 0.5$. pp = percentage points.}
\end{table*}

The Bayesian framework provides a substantial improvement over simple majority voting. Overall accuracy increases from 65.9\% to 82.9\% (+17.1 percentage points). Sensitivity improves dramatically from 58.3\% to 87.5\% (+29.2 percentage points), indicating that Bayesian weighting correctly recovers many true mergers that the majority vote misses. Specificity remains unchanged at 76.5\%, demonstrating that the framework maintains its performance on isolated galaxies while substantially improving merger detection.

\subsection{Per-Galaxy Classification Performance}
\label{sec:results_4_4}

Following L21, we assess classification performance using confident classifications ($p_M \geq 0.8$ or $p_M \leq 0.2$). Table~\ref{tab:confident} summarizes the results.

\begin{table*}
\caption{Confident Classification Results}
\label{tab:confident}
\centering
\begin{tabular}{lcccc}
\hline
Category & Total & Correct & Wrong & Purity \\
\hline
Confident Mergers ($p_M \geq 0.8$)  & 16 & 14 & 2 & 87.5\% \\
Confident Isolated ($p_M \leq 0.2$) & 14 & 11 & 3 & 78.6\% \\
Total Confident                     & 30 & 25 & 5 & 83.3\% \\
Uncertain ($0.2 < p_M < 0.8$)       & 11 & --- & --- & --- \\
\hline
\end{tabular}
\tablecomments{Following L21, we assess classification performance using confident classifications only ($p_M \geq 0.8$ or $p_M \leq 0.2$). The completeness—fraction of true mergers classified with $p_M \geq 0.8$—is 58.3\% (14/24).}
\end{table*}

\begin{figure*}
    \centering
    \includegraphics[width=\linewidth]{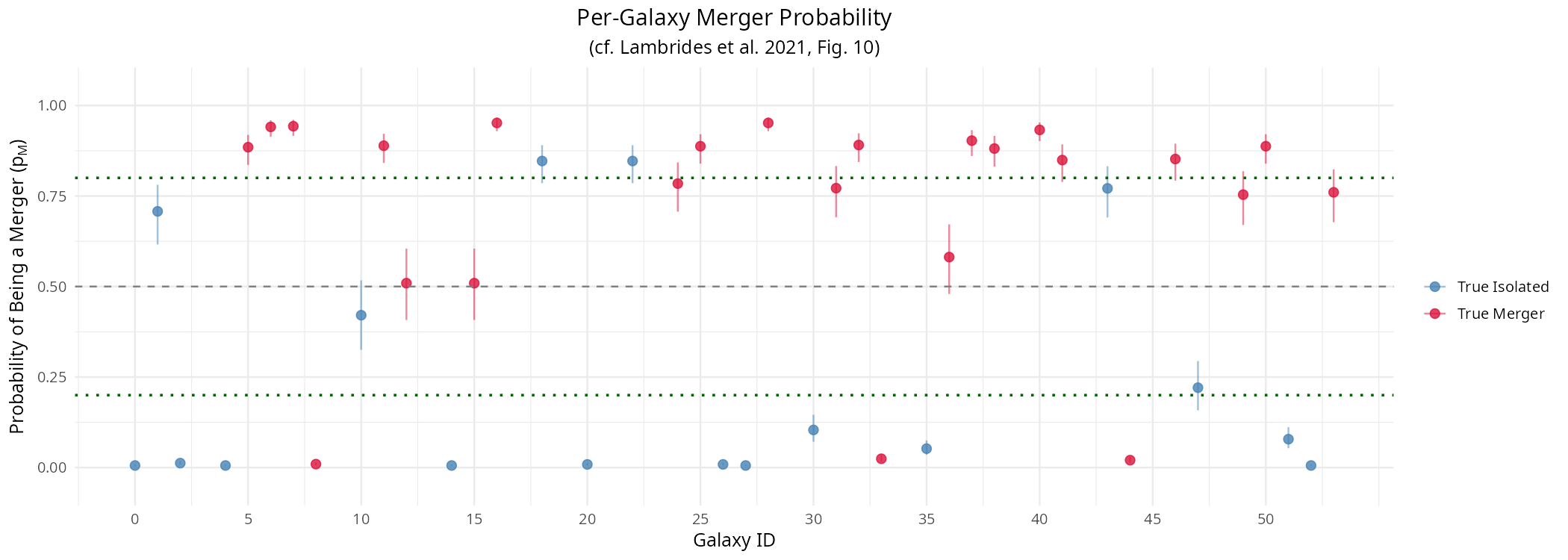}
    \caption{
    Per-galaxy merger probabilities $p_{\rm M}$ for the 41 mock galaxies.
    Confident mergers ($p_{\rm M} \ge 0.8$), confident isolated systems ($p_{\rm M} \le 0.2$), and uncertain cases are indicated, with misclassified galaxies highlighted.
    }
    \label{fig:pM_per_galaxy}
\end{figure*}

Of the 41 galaxies, 30 (73\%) receive confident classifications (Figure \ref{fig:pM_per_galaxy}): 16 as confident mergers ($p_M \geq 0.8$) and 14 as confident isolated ($p_M \leq 0.2$). The remaining 11 galaxies (27\%) fall in the uncertain range ($0.2 < p_M < 0.8$). This yields an accuracy of 83.3\% (25/30 confident classifications correct) with 5 misclassified galaxies (Table \ref{tab:misclassified}), closely matching the 85\% accuracy and 6 misclassifications reported by L21 for human classifiers. The completeness—the fraction of true mergers classified with $p_M \geq 0.8$—is 58.3\% (14/24).

\begin{table*}
\caption{Misclassified Galaxies}
\label{tab:misclassified}
\centering
\begin{tabular}{lcccc}
\hline
Galaxy & True Class & $p_M$  & Classification      & Error Type      \\
\hline
08    & Merger   & 0.009 & Confident Isolated & False Negative \\
33    & Merger   & 0.024 & Confident Isolated & False Negative \\
44    & Merger   & 0.021 & Confident Isolated & False Negative \\
18    & Isolated & 0.847 & Confident Merger   & False Positive \\
22    & Isolated & 0.847 & Confident Merger   & False Positive \\
\hline
\end{tabular}
\tablecomments{The three false negatives are true mergers with near-zero $p_M$ values, likely late-stage post-coalescence systems with faint merger signatures. The two false positives are isolated galaxies with morphological features the ensemble interpreted as merger evidence.}
\end{table*}

We examine how classification performance varies with redshift across the sample ($z = 1.08$--$3.55$). Completeness declines systematically with redshift: from 80\% at $z < 1.5$, to 71\% at $1.5 \leq z < 2.0$, 50\% at $2.0 \leq z < 2.5$, and 25\% at $z \geq 2.5$. Two of the three false negatives (galaxies 08 and 44) are at $z = 2.03$, consistent with the redshift threshold identified by L21, who found that removing post-mergers with $z > 2.0$ raises human completeness to 90\% and accuracy to 92\%. Applying the same cut to our sample raises accuracy to 88\% (23/26 confident classifications). The one remaining false negative (galaxy 33, $z = 1.56$) indicates that post-coalescence misclassification is not purely a redshift effect: morphological ambiguity from faded merger signatures can persist even at moderate redshift, as discussed for post-merger systems in Section~\ref{sec:disc_5_5}. 
Notably, the decline in completeness at high redshift manifests
primarily as an increased incidence of uncertain classifications
rather than confident misclassifications. The 11 uncertain galaxies
($0.2 < p_M < 0.8$) have a median redshift of
$z_{\rm med}=2.33$, compared with $z_{\rm med}=1.67$ for the
14 confidently correct mergers. Thus, higher-redshift systems are
preferentially assigned intermediate merger probabilities, consistent
with the progressive loss of detectable morphological evidence with
redshift. This behavior is consistent with the VLM ensemble appropriately hedging when morphological evidence is ambiguous, rather than committing to an incorrect classification.


Table \ref{tab:per_galaxy} presents the complete per-galaxy classification results, comparing majority-vote predictions with Bayesian-weighted probabilities for all 41 galaxies.

\begin{deluxetable*}{cccccccc}
\tablecaption{Per-Galaxy Classification Results\label{tab:per_galaxy}}
\tablehead{
  \colhead{Galaxy ID} & \colhead{True Class} & \colhead{MV Pred} &
  \colhead{MV Correct} & \colhead{$p_M$} & \colhead{Bayes Pred} &
  \colhead{Bayes Correct} & \colhead{Confidence}
}
\startdata
00 & Isolated & Isolated & Yes & 0.006 & Isolated & Yes & Confident \\
01 & Isolated & Merger & No & 0.708 & Merger & No & Uncertain \\
02 & Isolated & Isolated & Yes & 0.012 & Isolated & Yes & Confident \\
04 & Isolated & Isolated & Yes & 0.006 & Isolated & Yes & Confident \\
05 & Merger & Merger & Yes & 0.885 & Merger & Yes & Confident \\
06 & Merger & Merger & Yes & 0.941 & Merger & Yes & Confident \\
07 & Merger & Merger & Yes & 0.943 & Merger & Yes & Confident \\
08 & Merger & Isolated & No & 0.009 & Isolated & No & Confident* \\
10 & Isolated & Isolated & Yes & 0.422 & Isolated & Yes & Uncertain \\
11 & Merger & Merger & Yes & 0.889 & Merger & Yes & Confident \\
12 & Merger & Isolated & No & 0.510 & Merger & Yes & Uncertain \\
14 & Isolated & Isolated & Yes & 0.006 & Isolated & Yes & Confident \\
15 & Merger & Isolated & No & 0.510 & Merger & Yes & Uncertain \\
16 & Merger & Merger & Yes & 0.952 & Merger & Yes & Confident \\
18 & Isolated & Merger & No & 0.847 & Merger & No & Confident* \\
20 & Isolated & Isolated & Yes & 0.009 & Isolated & Yes & Confident \\
22 & Isolated & Merger & No & 0.847 & Merger & No & Confident* \\
24 & Merger & Isolated & No & 0.785 & Merger & Yes & Uncertain \\
25 & Merger & Isolated & No & 0.888 & Merger & Yes & Confident \\
26 & Isolated & Isolated & Yes & 0.009 & Isolated & Yes & Confident \\
27 & Isolated & Isolated & Yes & 0.006 & Isolated & Yes & Confident \\
28 & Merger & Merger & Yes & 0.952 & Merger & Yes & Confident \\
30 & Isolated & Isolated & Yes & 0.104 & Isolated & Yes & Confident \\
31 & Merger & Isolated & No & 0.772 & Merger & Yes & Uncertain \\
32 & Merger & Merger & Yes & 0.891 & Merger & Yes & Confident \\
33 & Merger & Isolated & No & 0.024 & Isolated & No & Confident* \\
35 & Isolated & Isolated & Yes & 0.052 & Isolated & Yes & Confident \\
36 & Merger & Isolated & No & 0.582 & Merger & Yes & Uncertain \\
37 & Merger & Merger & Yes & 0.903 & Merger & Yes & Confident \\
38 & Merger & Merger & Yes & 0.882 & Merger & Yes & Confident \\
40 & Merger & Merger & Yes & 0.933 & Merger & Yes & Confident \\
41 & Merger & Merger & Yes & 0.850 & Merger & Yes & Confident \\
43 & Isolated & Merger & No & 0.772 & Merger & No & Uncertain \\
44 & Merger & Isolated & No & 0.021 & Isolated & No & Confident* \\
46 & Merger & Merger & Yes & 0.853 & Merger & Yes & Confident \\
47 & Isolated & Isolated & Yes & 0.221 & Isolated & Yes & Uncertain \\
49 & Merger & Merger & Yes & 0.755 & Merger & Yes & Uncertain \\
50 & Merger & Isolated & No & 0.888 & Merger & Yes & Confident \\
51 & Isolated & Isolated & Yes & 0.079 & Isolated & Yes & Confident \\
52 & Isolated & Isolated & Yes & 0.006 & Isolated & Yes & Confident \\
53 & Merger & Merger & Yes & 0.761 & Merger & Yes & Uncertain \\
\enddata
\tablecomments{MV = majority vote; Bayes Pred uses the threshold $p_M > 0.5$ for mergers. Misclassified confident galaxies are marked with an asterisk. Galaxies 18 and 22 (isolated, $p_M \geq 0.8$) are false positives; 08, 33, 44 (mergers, $p_M \leq 0.2$) are false negatives.}
\end{deluxetable*}

\subsection{Comparison with Human Classifiers}
\label{sec:results_4_7}

Table \ref{tab:vlm_vs_humans} presents a direct comparison between our VLM ensemble results 
and the human classifier performance reported in L21, using identical 
confidence thresholds. While L21 sampled classifier accuracies as 
parameters, we fix accuracies to their calibration values 
(Section~\ref{sec:accuracies}); as shown in Section~\ref{sec:results_fixed_vs_sampled}, 
this methodological choice does not affect the estimated merger fraction 
but does influence the per-galaxy probability distributions.

\begin{table*}
\caption{Comparison of VLM Ensemble vs. Human Classifiers from L21}
\label{tab:vlm_vs_humans}
\centering
\begin{tabular}{lcc}
\hline
Metric & This Work (VLMs) & Lambrides et al. (Humans) \\
\hline
Number of galaxies               & 41          & 41          \\
Number of classifiers            & 15          & 14          \\
True merger fraction             & 59\%        & 59\%        \\
Estimated $f_{\rm M}$                  & 0.52 $\pm$ 0.09 & 0.52 $\pm$ 0.08 \\
$f_{\rm M}$ within 1$\sigma$ of truth  & Yes (0.66$\sigma$) & Yes     \\
Confidence threshold             & $p_M \geq 0.8$ / $\leq 0.2$ & $p_M \geq 0.8$ / $\leq 0.2$ \\
Accuracy (confident only)        & 83.3\%      & 85\%        \\
Misclassified galaxies           & 5           & 6           \\
False positives                  & 2           & 5           \\
False negatives                  & 3           & 1           \\
Completeness ($p_M \geq 0.8$)    & 58.3\%      & 85\%        \\
Uncertain classifications        & 11 (27\%)   & $\sim$6 (15\%) \\
\hline
\end{tabular}
\end{table*}

Both analyses use identical confidence thresholds and the same 41-galaxy VELA+SUNRISE mock galaxy sample, enabling direct comparison. The VLM ensemble achieves comparable accuracy with fewer false positives but more false negatives, reflecting a more conservative classification behavior.

The VLM ensemble achieves an accuracy of 83.3\%, closely matching the 85\% achieved by trained human classifiers, with 5 misclassified galaxies versus 6 for humans. Critically, the error profiles differ qualitatively: human classifiers produced 5 false positives (isolated galaxies confidently misclassified as mergers) and only 1 false negative, while the VLM ensemble produces 2 false positives and 3 false negatives. 

The VLM ensemble achieves lower completeness than human classifiers (58.3\% vs.\ 85\%), because its per-galaxy $p_M$ values are less polarized and more often fall into the uncertain region. However, the statistical significance of this difference is marginal (two-sided Fisher's test p = 0.11), and is discussed in Section \ref{sec:disc_5_4}.

\subsection{Fixed vs.\ Sampled Classifier Accuracies}
\label{sec:results_fixed_vs_sampled}

We tested an alternative model following L21 exactly, in which classifier accuracies $r_{M,i}$ and $r_{I,i}$ are sampled as parameters with informative Beta priors derived from calibration performance. Comparing the two approaches reveals systematic differences (Table~\ref{tab:fixed_vs_sampled}).

The inferred population merger fraction is nearly identical between the two treatments ($f_{\mathrm{M}} = 0.52 \pm 0.09$ for fixed accuracies vs.\ $f_{\mathrm{M}} = 0.51 \pm 0.08$ for sampled accuracies), so $f_{\mathrm{M}}$ itself is robust to this modeling choice. However, the per-galaxy posteriors differ: the sampled-accuracy model attains lower accuracy on confident classifications (74.4\% vs.\ 83.3\%), with 10 rather than 5 misclassified galaxies and many fewer uncertain cases (2 vs.\ 11). In particular, four true mergers move from the intermediate-probability (``uncertain'') regime under the fixed-accuracy model ($p_{\mathrm{M}} = 0.51$--$0.79$) to confidently isolated classifications under the sampled-accuracy model ($p_{\mathrm{M}} = 0.08$--$0.23$).
Overall classification accuracy is higher for the fixed-accuracy model (83.3\% vs.\ 74.4\%), while completeness is higher for the sampled-accuracy model (66.7\% vs.\ 58.3\%).

The sampled-accuracy model produces more polarized per-galaxy probability distributions, with fewer galaxies in the uncertain range ($0.2 < p_M < 0.8$): only 5\% of galaxies fall in the uncertain range compared to 27\% for the fixed-accuracy model. This behavior resembles the highly polarized distributions reported for human classifiers in L21, where $\sim$73\% of galaxies received $p_M$ values near 0 or 1. We discuss the origin of this behavior in Section~\ref{sec:discuss_fixed_vs_sampled}.

\subsection{Sensitivity to sampling variations}
\label{sec:sampling-uncertainty}

The primary claim of this paper is that VLM classifiers, when embedded 
in the Bayesian framework of L21, achieve classification accuracy 
statistically indistinguishable from that of human classifiers on this 
mock sample; the resampling analysis below quantifies the sampling 
uncertainty on the metrics underlying that comparison.

To assess how sensitive our conclusions are to the finite size of the
41-galaxy mock sample, we quantify sampling uncertainty using
leave-one-out jackknife resampling and non-parametric bootstrap
replicates of the per-galaxy classifications. These tests are applied to
the fixed-accuracy ensemble that underlies our main results; the
per-classifier sensitivities and specificities are held fixed at their
calibration values and are not resampled.

For the jackknife analysis, we remove each galaxy in turn and recompute
aggregate metrics on the remaining 40 systems. The maximum leave-one-out
shifts are modest: the Bayesian ensemble's overall accuracy changes by
at most 2.1 percentage points (from 82.9\% to 85.0\%), the sensitivity
and specificity by at most 3.8 and 4.8 percentage points, respectively,
and the accuracy on confident classifications by 2.9 percentage points;
the completeness changes by at most 2.5 percentage points. 

For the bootstrap analysis, we construct $B = 10{,}000$ resamples,
each containing 41 galaxies drawn independently from the original
41-galaxy sample with replacement; thus, a given galaxy may appear
multiple times in a resample, whereas others may not appear. We hold
the per-galaxy classifications and the fixed classifier accuracies
unchanged, and recompute the ensemble performance metrics for each
resample. Table~\ref{tab:uncertainty} reports the resulting 68\% and
95\% confidence intervals. For the Bayesian ensemble on the full sample
we obtain an accuracy of 82.9\% with 68\% CI $[78.0, 87.8]$ and 95\% CI
$[70.7, 92.7]$, sensitivity 87.5\% with 68\% CI $[80.8, 95.2]$, and
specificity 76.5\% with 68\% CI $[66.7, 86.7]$. For confident
classifications, the accuracy is 83.3\% with 68\% CI $[76.7, 90.3]$ and
95\% CI $[69.0, 96.4]$, and the completeness is 58.3\% with 68\% CI
$[48.0, 68.2]$ and 95\% CI $[38.1, 77.8]$. The number of confident
classifications varies from 24 to 35 across bootstrap realizations; we
propagate this variation rather than conditioning on exactly 30 confident
galaxies.

Table~\ref{tab:uncertainty} also reports the sampling uncertainty on 
the Bayes-vs-majority-vote comparison from Table~\ref{tab:mv_vs_bayes}. These differences 
confirm that L21's Bayesian framework is functioning as expected when 
applied to VLM classifiers, consistent with the original results for 
human classifiers in L21. However, this internal consistency check is not the primary goal of this work, which is to establish whether VLM-based classification, once embedded in this framework, achieves accuracy statistically indistinguishable from trained human classifiers on the same truth-known sample.

The paired bootstrap distributions show that Bayesian weighting
improves overall accuracy relative to majority voting by
$+17.1$~percentage points, with a 68\% CI of [12.2, 22.0] and a
95\% CI of [7.3, 29.3], and improves merger sensitivity by
$+29.2$~percentage points, with a 68\% CI of [20.0, 38.5] and a
95\% CI of [12.0, 48.1]. 
Exact McNemar tests \citep{McNemar1947} identify seven discordant
galaxies for both overall accuracy and merger sensitivity: all seven
are classified correctly by the Bayesian-weighted ensemble but
incorrectly by majority voting, while no galaxy is classified
correctly by majority voting but incorrectly by the Bayesian
ensemble. This corresponds to $p = 0.016$; there are no discordant
pairs among the isolated galaxies.

On the confident subset of 30 galaxies, the Bayesian ensemble achieves
83.3\% accuracy compared to 76.7\% for majority voting, a gain of 6.7
percentage points. However, only two galaxies are discordant between the
two methods, both favoring the Bayesian ensemble, and the exact McNemar
test yields $p = 0.50$; with this sample size, the improvement on
confident classifications is not statistically significant. Finally, we
confirm that the Bayesian decision threshold is stable: sweeping the
$p_M$ cut from 0.45 to 0.50 leaves the overall accuracy unchanged at
82.9\%, which only drops to 78.1\% at 0.55 and 75.6\% at 0.60--0.70;
only two galaxies (IDs~12 and~15) lie within 0.05 of the nominal
$p_M = 0.5$ threshold.

All resampling analyses were performed in \texttt{R} \citep{RCoreTeam2024}.

\begin{table}
\centering
\caption{Comparison of fixed-accuracy and sampled-accuracy models.}
\label{tab:fixed_vs_sampled}
\begin{tabular}{lcc}
\hline
\hline
Metric & Fixed & Sampled \\
\hline
Merger fraction ($f_M$) & $0.52 \pm 0.09$ & $0.51 \pm 0.08$ \\
Accuracy & 83.3\% & 74.4\% \\
Completeness & 58.3\% & 66.7\% \\
Misclassified galaxies & 5 & 10 \\
False positives & 2 & 4 \\
False negatives & 3 & 6 \\
Uncertain classifications & 11 (27\%) & 2 (5\%) \\
\hline
Mean $\Delta r_M$ (sampled $-$ fixed) & \multicolumn{2}{c}{$+0.047$} \\
Mean $\Delta r_I$ (sampled $-$ fixed) & \multicolumn{2}{c}{$+0.117$} \\
\hline
\end{tabular}
\tablecomments{Accuracy is computed on confident classifications only ($p_M \geq 0.8$ or $p_M \leq 0.2$); galaxies with intermediate probabilities are excluded. The fixed-accuracy model holds classifier accuracies constant at their calibration values; the sampled-accuracy model treats them as parameters with Beta priors following L21. $\Delta r_M$ and $\Delta r_I$ indicate the mean difference between sampled posterior means and fixed calibration values across all 15 classifiers.}
\end{table}

\begin{deluxetable*}{lccccc}
\tablecaption{{\bf Sampling Uncertainty on Aggregate Performance Metrics}\label{tab:uncertainty}}
\tablehead{
  \colhead{Metric} & \colhead{Value} & \colhead{Bootstrap $\sigma$} &
  \colhead{Max.\ LOO shift} & \colhead{68\% CI} & \colhead{95\% CI} \\
  \colhead{} & \colhead{(\%)} & \colhead{(pp)} & \colhead{(pp)} &
  \colhead{(\%)} & \colhead{(\%)}
}
\startdata
\cutinhead{Bayesian ensemble}
Accuracy (all 41) & $82.9$ & $5.9$ & $2.1$ & $[78.0,\,87.8]$ & $[70.7,\,92.7]$ \\
Sensitivity $r_M$ & $87.5$ & $6.9$ & $3.8$ & $[80.8,\,95.2]$ & $[72.7,\,100.0]$ \\
Specificity $r_I$ & $76.5$ & $10.4$ & $4.8$ & $[66.7,\,86.7]$ & $[55.0,\,94.4]$ \\
Accuracy (confident only) & $83.3$ & $6.9$ & $2.9$ & $[76.7,\,90.3]$ & $[69.0,\,96.4]$ \\
Completeness ($p_M\geq0.8$) & $58.3$ & $10.1$ & $2.5$ & $[48.0,\,68.2]$ & $[38.1,\,77.8]$ \\
\cutinhead{Majority vote}
Accuracy (all 41) & $65.9$ & $7.4$ & $1.6$ & $[58.5,\,73.2]$ & $[51.2,\,80.5]$ \\
Sensitivity $r_M$ & $58.3$ & $10.2$ & $2.5$ & $[48.0,\,68.2]$ & $[38.1,\,77.8]$ \\
Specificity $r_I$ & $76.5$ & $10.4$ & $4.8$ & $[66.7,\,86.7]$ & $[55.0,\,94.4]$ \\
\cutinhead{Bayesian $-$ majority vote (paired)}
$\Delta$ Accuracy (all 41) & $17.1$ & $5.9$ & $2.1$ & $[12.2,\,22.0]$ & $[7.3,\,29.3]$ \\
$\Delta$ Sensitivity $r_M$ & $29.2$ & $9.4$ & $3.1$ & $[20.0,\,38.5]$ & $[12.0,\,48.1]$ \\
$\Delta$ Specificity $r_I$ & $0.0$ & $0.0$ & $0.0$ & $[0.0,\,0.0]$ & $[0.0,\,0.0]$ \\
\enddata
\tablecomments{Bootstrap statistics use $B=10{,}000$ resamples of the 41 galaxies drawn with replacement. ``Max.\ LOO shift'' is the largest change in the metric produced by removing any single galaxy (leave-one-out jackknife, 41 replicates); pp = percentage points. The paired differences in the final block are evaluated within each bootstrap resample, since the two methods are scored on the same galaxies. The specificity difference is identically zero: the two methods return identical predictions for all 17 isolated galaxies, so no confidence interval is defined. Exact McNemar tests on the paired differences give $p=0.016$ for both the accuracy and the sensitivity gain (7 discordant galaxies, all favoring the Bayesian ensemble).}
\end{deluxetable*}

\section{Discussion}
\label{sec:discussion}

\subsection{VLMs as Viable Alternatives to Human Classifiers}
\label{sec:disc_5_1}

Our results demonstrate that, when embedded in the Bayesian framework
of L21 and applied to the identical truth-known mock sample, an ensemble
of Vision-Language Models attains galaxy-merger classification accuracy
that is statistically indistinguishable from that of trained human
experts on this 41-galaxy test. The VLM ensemble achieves 83.3\%
accuracy on confident classifications with 5 misclassified galaxies,
compared to 85\% accuracy and 6 misclassifications for the human
classifiers of L21. The 95\% bootstrap confidence interval on the VLM
accuracy covers 69.0--96.4\%, which fully encompasses the human
benchmark; a larger sample would be required to detect or rule out any
small residual performance difference. Within this mock domain, the
near-identical performance suggests that current small VLMs have reached
a level of visual pattern recognition sufficient for galaxy morphology
classification, without any task-specific training. This comparison is
methodologically robust: both studies use the same VELA+SUNRISE mock
galaxy sample, the same Bayesian statistical framework, the same
confidence thresholds, and comparable numbers of classifiers (15 VLMs
vs.\ 14 humans).



The Bayesian framework of L21 proves equally effective for combining
VLM classifications as for human classifications. The improvement from
majority voting to Bayesian-weighted classification (+17.1 percentage
points in overall accuracy, +29.2 percentage points in sensitivity;
McNemar $p = 0.016$ for both; see Section~\ref{sec:sampling-uncertainty})
demonstrates the importance of accounting for individual classifier
accuracies, particularly given the substantial variation in performance
across different VLM configurations.

The error profiles of VLM and human classifiers differ qualitatively. 
Human classifiers produced 5 false positives and only 1 false negative, 
while the VLM ensemble produces 2 false positives and 3 false negatives 
with our fixed-accuracy approach. When we apply the sampled-accuracy 
method of L21 to the VLM ensemble (Section~\ref{sec:results_fixed_vs_sampled}), 
the error counts increase to 4 false positives and 6 false negatives, 
but the asymmetry persists: VLMs produce more false negatives than 
false positives, while humans show the opposite pattern. This suggests 
a qualitative difference in classification behavior, with VLMs being 
more conservative (missing mergers) and humans being more liberal 
(over-identifying mergers),  though the current sample is too small to establish this trend as statistically significant (see Section~\ref{sec:disc_5_4}).

\subsection{Impact of Prompt Engineering and Ensemble Diversity}
\label{sec:disc_5_3}

Our results confirm that prompt design has a substantial, systematic effect on individual VLM classifier performance. The diversity of prompt strategies is a feature rather than a liability of the ensemble. The four strategies produce systematically different classification behaviors: open prompts yield mean $r_M = 65.6$\% and mean $r_I = 37.5$\% (merger-biased), optimal prompts achieve mean $r_M = 69.8$\% and mean $r_I = 51.0$\% (balanced), balanced prompts produce mean $r_M = 44.4$\% and mean $r_I = 61.5$\% (isolated-biased), and strict prompts yield mean $r_M = 36.5$\% and mean $r_I = 67.3$\% (isolated-biased). 
These prompt strategies produce classifiers spanning a broad range of
operating points in the $r_{\rm M}$--$r_{\rm I}$ plane, a property
that, together with sufficient individual discriminative power,
appears important for effective Bayesian combination
(see below).

As a cross-check, we compared the 12-classifier ensemble (omitting the balanced prompts) to the full 15-classifier ensemble and found no material change in the recovered merger fraction, overall accuracy, or completeness; the small differences in confident-classification accuracy (at the level of a single galaxy out of 41) are consistent with small-number fluctuations. This indicates that the Bayesian weighting scheme correctly down-weights poorly calibrated classifiers and up-weights well-calibrated ones. The balanced-prompt classifiers, despite their individually moderate performance, contribute marginally to the ensemble through the diversity of their error patterns. 

We further tested a more restrictive subset, retaining only the
open- and strict-prompt classifiers (8 of 15 total). This produces a substantial degradation: overall accuracy after Bayesian correction falls to 73.2\% (vs. 82.9\% for the full ensemble), no galaxy
reaches the confident-merger threshold ($p_M \geq 0.8$; maximum
inferred $p_M = 0.667$), completeness falls to 0\%, and the
recovered merger fraction falls outside the 1$\sigma$ interval
of the true value. Using Youden's $J$ statistic \citep{Youden1950}
($J = r_M + r_I - 1$), for which larger values indicate greater
joint discriminative power and $J=0$ corresponds to chance-level
discrimination, this 8-classifier subset has a maximum $J = 0.21$
and mean $J = 0.035$, compared to a maximum $J = 0.30$ and mean
$J = 0.085$ for the full 15-classifier ensemble; 
the two highest-J classifiers (IDs 3 and 7, both using the "optimal" prompt) are excluded from this restricted subset.
Open and strict prompts push sensitivity and specificity in opposite
directions simultaneously, so individual classifiers in this subset
lie close to the chance diagonal, $r_M + r_I = 1$ (equivalently,
$J=0$), even though the subset spans a wide numerical range in each
quantity separately. 
Since this subset differs from the full ensemble in prompt diversity,
individual discriminative power, and ensemble size, the test does not
disentangle their effects. For this ensemble and sample, it shows
that variation in $r_{\rm M}$ and $r_{\rm I}$ alone is insufficient;
robust performance appears to depend on a combination of classifier
diversity and sufficiently strong individual discriminative power.

To assess whether the optimal-prompt classifiers alone could
reproduce the full ensemble's performance, we tested the subset
of four optimal-prompt configurations. Three of these four classifiers have similar sensitivity and specificity. This subset yields
$f_{\rm M}=0.437^{+0.127}_{-0.132}$, with 73.3\% accuracy
on confident classifications and 0\% completeness.
This subset does not reproduce the full ensemble's performance,
although the comparison does not disentangle the effects of
diversity in operating points (sensitivity/specificity) and
ensemble size.

Beyond testing the effect of prompt diversity, we also examined
whether a small, favorably selected subset of classifiers could
approximate the performance of the full ensemble. 
We tested a specific subset of five classifiers chosen for their
high individual discriminative power, as measured by Youden's
$J = R_M + R_I - 1$ (Classifiers 1, 2, 3, 4, and 7; all Gemma-4
E2B/E4B variants). This cherry-picked subset recovered
$f_M = 0.517^{+0.105}_{-0.104}$, consistent with the true value
($f_M = 0.585$) at the $0.66\sigma$ level, comparable to the full
15-classifier ensemble result. 
However, the accuracy on confident classifications
($p_M > 0.8$ or $p_M < 0.2$) is 79.4\% for this subset, compared
with 83.3\% for the full ensemble and 85\% for the human-classifier
benchmark of L21. Its completeness is 62.5\%,
higher than that of the full ensemble (58.3\%) but lower than the
human benchmark (85\%). Given the small sample and the associated
sampling uncertainty, these apparent differences should not be
interpreted as statistically significant. This
deliberately selected, high-$J$ subset does not establish a general
minimum ensemble size, since an arbitrary or less favorable subset
of five classifiers could perform substantially worse.

The Gemma-4 family responds predictably and proportionally to prompt strictness, making it the most controllable family for calibration purposes. Qwen3-VL is highly instruction-following but exhibits extreme swings across prompt variants ($r_M$–$r_I$ dynamic range of 71 percentage points). In contrast, Qwen2.5‑VL exhibits minimal responsiveness to prompt framing: a balanced prompt produced merger and non‑merger recalls essentially identical to the optimal variant, reinforcing the picture that this model’s behavior is largely governed by internal priors rather than subtle changes in instruction wording.

We note that because each prompt family was independently tuned
per architecture (Section~\ref{sec:prompts}), these per-family
behavioral characterizations describe specific model-prompt
configurations rather than intrinsic architectural properties; no
claim is made about the relative capability of one architecture
versus another.

The overlap between $r_M$ and $r_I$ posteriors visible for
classifiers 4 and 6 in Figure~\ref{fig:rM_rI_distribution}
arises because their sensitivity and specificity estimates are
numerically close to each other relative to the sampling
uncertainty of our calibration sample: the differences
$|r_M - r_I|$ for these classifiers (0.13 and 0.08, respectively)
are comparable to or smaller than the binomial uncertainty on
each estimate, making the two quantities statistically
indistinguishable. The same reasoning applies to classifier 9 (Qwen2.5-VL, open
prompt; $r_M = 0.375$, $r_I = 0.538$), whose apparent separation
from classifiers 4 and 6 is likewise not statistically significant
given binomial sampling uncertainty at this sample size.
Classifiers whose $r_M$ and $r_I$ posteriors
overlap in this way contribute comparatively little independent
information to the ensemble, and are naturally downweighted by
the Bayesian framework relative to more discriminative
classifiers.

\subsection{The Completeness–Purity Trade-off}
\label{sec:disc_5_4}

The primary difference between VLM and human classification is completeness: 58.3\% for VLMs versus 85\% for humans. However, a two-sided Fisher's exact test comparing these proportions (14/24 vs.\ 20/24) yields $p = 0.11$, indicating that the difference is not statistically significant at conventional levels, and larger samples are needed to confirm whether this gap is genuine. The corresponding 95\% bootstrap interval for VLM completeness is
38--78\% (Section~\ref{sec:sampling-uncertainty}), illustrating the
substantial sampling uncertainty at the current sample size.
The lower VLM completeness arises because VLMs place more galaxies in the uncertain category (27\% vs.\ $\sim$15\%), reflecting more moderate $p_M$ values from the 15-classifier ensemble---though this difference is also not statistically significant given the sample size.

The three false negatives (08, 33, 44) are true mergers with near-zero inferred $p_M$ values (0.009, 0.024, and 0.021), suggesting that the ensemble was driven toward the wrong answer by a majority of classifiers, rather than being merely uncertain. Visual inspection of these galaxies reveals common morphological features that challenge VLM classification. These galaxies are late-stage post-coalescence systems where merger signatures are faint, consistent with the known post-merger classification challenge documented in L21.

The two false positives (18, 22) are isolated galaxies with $p_M$ values just above the confident threshold (0.847 for both), indicating they sit near the decision boundary. The apparent difference in false positive rates (2/17 for VLMs vs. 5/17 for humans) is not statistically significant with these sample sizes. Similarly, the difference in false negative rates (3/24 vs. 1/24) cannot be distinguished from statistical fluctuation. Consequently, claims about differing failure modes between VLM and human classifiers must remain tentative until validated on larger samples.

For applications where false positives are costly (e.g., spectroscopic follow-up of merger candidates), the VLM ensemble's low absolute false positive count (2 galaxies) may be acceptable. For applications requiring high completeness, the $p_M$ threshold for confident mergers can be lowered, or the ensemble can be augmented with additional high-sensitivity classifiers.

To quantify this trade-off, we recompute completeness and purity
across a range of $p_M$ thresholds using the full ensemble. Lowering
the threshold from 0.8 to 0.5 increases completeness from 58.3\% to
87.5\%, while merger purity changes only modestly (87.5\% to
84.0\%), indicating a favorable completeness--purity trade-off for
applications that prioritize completeness. We emphasize that this
threshold-dependent exploration is distinct from our primary
comparison with L21 in Section~\ref{sec:results},
which is fixed at the $p_M \geq 0.8$ threshold used in their study.

\subsection{Fixed vs.\ Sampled Classifier Accuracies}
\label{sec:discuss_fixed_vs_sampled}

Section~\ref{sec:results_fixed_vs_sampled} presented results from the sampled-accuracy model of L21, in
which classifier accuracies ($r_M$, $r_I$) are inferred jointly with the
true galaxy classes rather than fixed at calibration values. As in L21,
the sampled model produces overconfident classifications: nearly all
galaxies receive $p_M$ values near 0 or 1, with only 5\% of galaxies
remaining uncertain compared to 27\% in the fixed model. This overconfidence
arises from a feedback loop in the joint inference: galaxies with polarized
votes are confidently assigned a true class matching the majority, which
inflates the inferred accuracies of classifiers who agreed with that
majority. Higher inferred accuracies in turn drive more extreme probability
assignments, reinforcing the cycle. This feedback loop is amplified by the coherence of our ensemble: when
most classifiers agree on a galaxy, the sampled model infers true
classes with high confidence, maximizing the credit assigned to the
agreeing classifiers and therefore inflating their inferred accuracies, which in turn drives overconfident per‑galaxy probability assignments.

We therefore recommend the fixed-accuracy approach for applications where
the calibration sample is approximately balanced between classes. Fixing
accuracies to calibration values anchors the inference to an external
measurement, breaks the $f_M$--accuracy degeneracy, and produces
better-calibrated per-galaxy uncertainties. The sampled-accuracy approach
may be appropriate when there is reason to believe classifier performance
differs systematically between the calibration and inference
samples---for example, if real galaxy images have substantially different
noise properties or morphological characteristics than the synthetic
calibration images---but users should be aware that it may produce
overconfident classifications.

We note that the comparison in Table~\ref{tab:vlm_vs_humans} is
not perfectly matched: it presents the VLM ensemble under our
fixed-accuracy treatment alongside the human classifiers of L21
under their published sampled-accuracy treatment, since L21 did
not report the classifier-level vote data needed to reanalyze
their sample under the fixed-accuracy model. Our preference for
the fixed-accuracy treatment is nonetheless methodologically
motivated rather than a selection of the more favorable result,
given the overconfidence pathology of the sampled model
described above. We also verified that the differences between the two treatments are
not statistically significant. A Fisher's exact test comparing
confident-classification outcomes for the fixed-accuracy and
sampled-accuracy models, respectively (25/5 versus 29/10
correct/wrong), yields $p \approx 0.56$. A corresponding test of
completeness, comparing 14/10 versus 16/8 identified/missed true
mergers for the fixed-accuracy and sampled-accuracy models,
respectively, yields $p \approx 0.77$.
Thus, although the sampled-accuracy model yields lower observed
confident-classification accuracy and higher observed completeness
in this sample, the present data do not establish that these
differences reflect a systematic trade-off between the two accuracy
treatments.

\subsection{Challenges: Post-merger Systems}
\label{sec:disc_5_5}

A recurring challenge across all model-prompt combinations is the classification of post-merger systems: galaxies that have coalesced into a single morphological object but retain internal signatures of a recent interaction. These systems represent a structurally ambiguous category where even human classifiers disagree, as documented in L21, who found that all but one of their misclassified galaxies are post-mergers, with accuracy dropping significantly at $z > 2.0$ where merger features have faded. The three VLM false negatives (08, 33, 44), with $p_M$ values of 0.009, 0.024, and 0.021 respectively, are consistent with this known challenge; their near-zero probabilities suggest that the ensemble of VLM classifiers, like human experts, is systematically misled by the morphological ambiguity of late-stage coalescences.

\subsection{Practical Advantages and Limitations of VLM Classification}
\label{sec:disc_5_6}

VLM-based classification offers several practical advantages over human classification that are directly relevant to large-scale galaxy morphology studies. VLMs can process thousands of galaxy images per hour without fatigue or inter-session variability. Under fixed model weights and prompt, outputs are reproducible, enabling fully auditable analyses. Prompt engineering allows fine control over the $r_M$–$r_I$ trade-off in a measurable and systematic way that is not possible with human classifiers. Finally, VLMs produce structured outputs with explicit evidence lists and confidence scores, facilitating post-hoc error analysis. In this proof-of-concept we deliberately restrict ourselves to small, locally runnable open-weight VLMs, so the results reported here should be interpreted as a conservative lower bound; larger or more capable future models are likely to improve both individual classifier performance and ensemble completeness.

Several important limitations must be noted. First, our validation is performed on mock galaxy images from a single simulation suite (VELA+SUNRISE); if real galaxy morphologies differ systematically from the simulations — due to higher dust content, different merger timescales, or redshift-dependent surface brightness effects — VLM performance may degrade. L21 note the same concern for human classifiers. However, the classifier biases measured here reflect not only the image properties but also the intrinsic interpretation styles of each model—whether a classifier is conservative or liberal in identifying merger features. These interpretation biases are driven by model architecture and prompt design rather than image-specific characteristics, suggesting they may be more stable across domains than the absolute accuracy values. If the relative biases (e.g., which classifier is more sensitive, which is more specific) are preserved when moving to a different image domain, the Bayesian framework may still provide partial correction benefits even without full recalibration. Nevertheless, for quantitatively reliable merger fraction estimates, recalibration on domain-matched mock images remains the recommended practice. Observational validation should likewise rely on independent and
complementary indicators of merger activity (e.g., spectroscopic
close pairs, tidal features, spatially resolved kinematic
disturbances, or independently identified late-stage mergers)
rather than on a single definitive classification
\citep{Desmons2023}. Human classifications can also serve as an
external benchmark when their selection effects and uncertainties
are acknowledged.

A further limitation is that our validation sample is restricted to
a single survey depth and a comparatively narrow range of halo
virial masses ($\sim$1--20 $\times$ 10$^{11}\, M_\odot$ at $z=2$;
Ceverino et al. 2014), inherited from the VELA simulation suite
underlying both this work and L21, and does not probe a range of
surface brightness limits. The mock images are specifically designed to simulate
HST/ACS+WFC3 observations of the GOODS-S field at 3D-HST depth and
filter coverage \citep{Simons2019}, and they incorporate a realistic
noise model for that observational configuration. Performance on
shallower surveys, at significantly different depths, or for
galaxies at stellar/halo masses outside this range may therefore
differ from the results reported here. Mass and surface brightness completeness effects are known to
influence merger identification rates in flux-limited samples
(e.g., Conselice 2014). While these are not quantified in the
present work, we identify their characterization, using mock images
spanning a range of survey depths and a broader range of halo and
stellar masses, as an important direction for future work.

\subsection{Implications for Large-Scale Surveys}
\label{sec:disc_5_7}

The proof-of-concept presented here is directly applicable to ongoing and planned imaging surveys. A pre-calibrated VLM ensemble can be deployed on any new survey field without additional human labeling: each model-prompt combination acts as a virtual classifier with a known ($r_M$, $r_I$) pair, and the full classification pipeline can be re-run identically whenever the mock calibration set or priors are updated — producing versioned, auditable merger catalogs that are not possible with frozen human vote collections.

The Roman Space Telescope \citep{Spergel2015} represents the most demanding future case. Its projected wide-area near-infrared imaging footprint, expected to yield morphological measurements for hundreds of millions of galaxies across cosmic time, makes human-based per-classifier calibration across the full survey footprint effectively impossible. A VLM ensemble integrated into a Roman data pipeline would produce per-galaxy $p_M$ values with a fully characterized and reproducible selection function, shifting the primary bottleneck from human throughput to the fidelity of the simulation calibration set. This is a challenge that will improve progressively with the next generation of hydrodynamical simulations. Large-volume surveys from Euclid and the Vera Rubin
Observatory likewise stand to benefit from scalable, reproducible morphological
classification using VLM ensembles.

\section{Conclusions}
\label{sec:conclusions}

We have demonstrated that an ensemble of Vision-Language Models, combined using the Bayesian statistical framework of L21, can classify galaxy merger status with accuracy comparable to trained human experts. Our principal findings are as follows.

The VLM ensemble achieves an accuracy of 83.3\% on confident classifications ($p_M \geq 0.8$ or $p_M \leq 0.2$) using identical methodology and thresholds to L21, closely matching the 85\% achieved by human classifiers. The VLM ensemble produces 5 misclassifications, comparable to the 6 reported for human classifiers. VLMs place more galaxies in the uncertain category (27\% vs.\ $\sim$15\%) and produce fewer false positives (2 vs.\ 5), which may suggest less overconfident classification behavior---however, these differences are not statistically significant given the sample size of 41 galaxies and warrant further investigation with larger samples.

The estimated merger fraction ($f_{\rm M} = 0.52 \pm 0.09$) is within $0.66\sigma$ of the true value (0.585), demonstrating successful recovery of population-level merger statistics, with excellent MCMC convergence ($\hat{R} = 1.002$, $n\_{\rm eff} > 1600$). Bayesian weighting improves overall classification accuracy by +17.1 percentage points over simple majority voting, with sensitivity improving by +29.2 percentage points. 
We find that fixing classifier accuracies to their calibration values, 
rather than sampling them as parameters following L21, produces 
better-calibrated per-galaxy probabilities and avoids a degeneracy 
between the inferred merger fraction and classifier accuracies. 
This methodological refinement yields 83.3\% accuracy compared to 
74.4\% with the sampled-accuracy approach.
Prompt engineering significantly impacts individual classifier performance, with open prompts producing merger-biased classifiers and strict prompts producing isolated-biased classifiers; 
robust ensemble performance appears to depend on a combination of
classifier diversity and sufficiently strong individual
discriminative power.
Using 15 VLM classifiers, directly comparable in number to the 14 human classifiers in L21, enables a robust apples-to-apples comparison demonstrating that VLMs achieve classification accuracy
statistically indistinguishable from that of trained human
classifiers on this 41-galaxy mock sample.



Within this mock domain, these results establish small, locally runnable VLM ensembles as practical
alternatives to human classifiers for galaxy morphology studies, offering comparable accuracy, reduced
false positives, computational scalability, and full reproducibility under fixed model checkpoints.
The primary limitation is lower completeness (58.3\% vs.\ 85\%), which can be addressed through threshold
adjustment, ensemble augmentation with additional high-sensitivity classifiers, or hybrid human–VLM
approaches, depending on application requirements. A natural next step is to develop and validate this framework for application to large
observational samples from HST legacy surveys and JWST, with appropriate recalibration on domain-matched
mock images and comparison with independent and complementary observational
merger indicators (e.g., spectroscopic close pairs, tidal features, spatially resolved kinematic disturbances, or independently identified late-stage mergers), which probe
different merger stages with distinct selection effects
\citep{Desmons2023}.

Longer term, larger locally deployable open-weight VLMs offer substantially richer visual
representations while remaining fully reproducible under fixed weights. Using locally hosted
models with frozen checkpoints is essential for scientific reproducibility, since API-accessible
frontier models can be updated, fine-tuned, or deprecated without notice, making it impossible
to guarantee that classifications or merger-fraction inferences can be exactly reproduced at a
later date.
Improved individual model capability is expected to translate into higher per-classifier accuracies within the Bayesian ensemble framework, making the testing of larger open-weight architectures a natural next step toward closing the completeness gap identified here. Beyond merger identification, the same VLM-based framework could be extended
to other morphology-driven tasks such as strong and weak gravitational lens
identification in large imaging surveys, where reproducible visual
classification is equally critical. 
The present proof-of-concept is validated on the 41-galaxy VELA+SUNRISE mock sample of L21, which was deliberately adopted to enable a direct, controlled comparison with the human classifier benchmark established in that work. A natural and prioritized next step is to extend this validation to larger simulation-based mock catalogs --- for example, IllustrisTNG-based mock images at HST and JWST depth --- which would allow a more statistically robust assessment of the framework's performance across a wider range of merger types, mass ratios, redshifts, and surface brightness regimes. 
We also plan to explore alternative sources of ensemble diversity,
such as multi-agent classifier interaction, in future work.
We emphasize that any such extension should use truth-known simulation labels rather than human-consensus classifications (e.g., Galaxy Zoo), since the framework is specifically designed to be independent of human labeling bias; validating against human labels would reintroduce the systematic biases the method is designed to circumvent. Human classifications can nonetheless serve as a useful external
benchmark when their uncertainties and selection effects are
acknowledged, without being treated as definitive ground truth.

\section*{Acknowledgments}

We thank Erini Lambrides for providing the mock galaxy images used in this work, and the original version of the MCMC Bayesian analysis code.
We acknowledge extensive use of Claude (Anthropic) for code development, 
statistical analysis, and manuscript preparation: Claude Opus 4.5 through 
Johns Hopkins University's HopGPT interface (v2.6.0), and Perplexity, primarily using Claude Sonnet 4.6. 
MC wishes to thank Stephan McCandliss for helpful discussions. We thank the anonymous referee for a careful and constructive report, whose suggestions substantially strengthened the statistical analysis and the presentation of this work.


\bibliographystyle{aasjournal}
\bibliography{mergers}

\appendix

\section{Use of Large language models}\label{sec:appendix_llm_use}

Beyond the VLM-based galaxy classification that is the subject of this study, 
this work was developed through extensive collaboration with Claude (Anthropic). 
Claude Opus~4.5 was accessed through Johns Hopkins University's HopGPT interface 
(v2.6.0) for extended analytical discussions, code development, and manuscript 
drafting. Perplexity, primarily using Claude Sonnet~4.6 with occasional 
use of its ``Best'' mode, assisted with literature 
searches and quick queries. The AI assistants contributed to multiple aspects 
of this project:

Code development. A new R implementation of the L21 Bayesian framework, including the Stan MCMC model, data ingestion pipelines, statistical analysis functions, and visualization code, was developed iteratively through dialogue with Claude. The authors provided specifications, reviewed outputs, identified bugs, and guided refinements.

Statistical analysis. Claude Opus 4.5 (and later versions up to 5.5 for the revised version) assisted with interpreting MCMC diagnostics, computing confidence intervals, performing Fisher's exact tests for significance assessment, and ensuring that statistical claims were appropriately caveated given the small sample sizes.

Manuscript preparation. Drafts of text, tables, and figure captions were generated and refined through iterative dialogue. The authors provided domain expertise, verified scientific accuracy, identified logical inconsistencies, and made editorial decisions about framing and emphasis.

Critical review. Claude served as a sounding board for methodological choices, flagged potential over-interpretations of small-number statistics, and helped ensure consistency between numerical results and textual claims throughout the manuscript.

The interaction model was collaborative: the authors posed questions, provided data and context, evaluated AI-generated outputs, and made all final decisions about scientific content. The AI assistant functioned as a highly capable research assistant rather than an autonomous agent—generating candidate text and code that was then critically evaluated, corrected, and approved by the human authors. All scientific conclusions and interpretations represent the judgment of the authors, who take full responsibility for the content of this work.

This approach is in the spirit of \citet{Schwartz2026}, who recently demonstrated that an AI assistant can contribute substantively to original physics research under close human supervision. Our workflow contrasts with
fully automated research pipelines such as ``The AI Scientist''
\citep{AIScientist2026}, which aim to automate the entire scientific
lifecycle from idea generation to manuscript production. In this work, AI
tools function strictly as collaborators and accelerators: the authors
designed the research questions, executed all code, validated every result,
and made all scientific and editorial decisions, with the AI assistants
providing candidate code and text that were then critically reviewed and
edited.

\section{VLM Implementation and Inference Details}
\label{sec:appendix_impl}

We employ four locally runnable vision--language model families in this work.
The Gemma-4 E2B and E4B configurations denote the 2B and 4B effective-parameter
on-device models, respectively; these models use per-layer embeddings (PLE) to
maximize parameter efficiency, so the total number of loaded weights is larger
than the quoted effective parameter count. All models are used in 4-bit GGUF
quantized form for efficient local inference. The specific GGUF model files are:
\begin{itemize}
\item \texttt{gemma-4-E2B-it-UD-Q4\_K\_XL.gguf}
\item \texttt{gemma-4-E4B-it-Q4\_K\_M.gguf}
\item \texttt{Qwen\_Qwen2.5-VL-7B-Instruct-Q4\_K\_M.gguf}
\item \texttt{Qwen3-VL-4B-Instruct-Q4\_K\_M.gguf}
\end{itemize}
For multimodal inference we additionally load the corresponding vision
projection files: \texttt{mmproj-F16.gguf} for the Gemma-4 E2B/E4B and
Qwen3-VL-4B models, and \texttt{mmproj-Qwen\_Qwen2.5-VL-7B-Instruct-f16.gguf}
for Qwen2.5-VL-7B. All model and projector GGUF files were obtained from
publicly available Hugging Face model hubs.


Qwen2.5-VL is evaluated with three prompt strategies (open, optimal, strict), while the other families use four (balanced, open, optimal, strict), yielding 15 classifier configurations in total. All vision projector (mmproj) components run at FP16 precision.

All models were run using \texttt{llama.cpp} (version 8857, build \texttt{a6cc43c28}), compiled for CPU-only operation on a Linux x86\_64 machine with GNU 13.3.0. The server was launched with \texttt{--ctx-size 8192} and \texttt{--n-predict 2048}; the generation limit was essential to prevent silent output truncation in verbose prompt variants, which otherwise caused JSON parse errors on galaxies that elicited long reasoning chains. This configuration represents a conservative lower bound on performance that would be achievable with GPU-accelerated inference or larger locally deployable models.

\section{Example VLM Prompt and Response for a Merger}
\label{sec:appendix_example_merger}

\begin{figure*}
\centering
\includegraphics[width=0.41\textwidth]{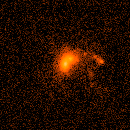}
\includegraphics[width=0.45\textwidth]{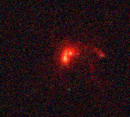}
\caption{
Example mock galaxy image (galaxy 07).
Left: WFC3/IR F160W image (rest-frame optical at $z \sim 1$--2) shown as a grayscale or heat-map intensity map.
Right: RGB composite combining F160W, F775W, and F435W, highlighting clumpy, asymmetric star formation.
This is the same system used in the example VLM classification in Appendix~\ref{sec:appendix_example_merger}, where both Qwen2.5-VL and Gemma-4 E4B assign the primary label \texttt{disturbance\_major}, collapsing to the binary \texttt{merging} class.
}
\label{fig:r07_example}
\end{figure*}

This appendix illustrates the full classification output for a clear merger system (galaxy 07) using the Qwen2.5-VL model with the optimal prompt (Section~\ref{sec:prompts}). The input is the two-panel postage stamp (F160W heatmap and RGB composite), and the model is queried via a local \texttt{llama.cpp}-compatible API.

\subsection{Gemma-4 E4B, Optimal Prompt}
\begin{spverbatim}
You are an expert extragalactic astronomer.

The galaxy at the center of the image is the target and it is at z between 1 and 2,
so you are seeing rest-frame optical structure.

Your task is to visually classify the system following Lambrides et al. (2021).

IMAGE INPUTS:
You are shown TWO images of the same galaxy:

1. Near-IR image (WFC3-IR, heat colormap): Rest-frame optical at z=1-2. Displayed
   using a DS9 heat colormap where black/dark = faint emission and
   red/orange/yellow/white = bright emission. Most sensitive for detecting faint,
   low surface brightness features like tidal tails, shells, and plumes.
   Use this as the PRIMARY image for identifying merger signatures.

2. RGB composite (Near-IR + optical WFC3-UVIS): The blue channel shows rest-frame
   UV at z=1-2, tracing young star formation. Use this to identify:
   - Blue star-forming clumps (normal in spiral arms)
   - Whether clumpy structure follows a spiral pattern (normal) or is chaotic (potential merger)


REQUIRED REASONING STEPS — follow these in order:

  Step 1: Describe the near-IR morphology. Note any elongation, asymmetry,
          clumpiness, or extensions, however subtle.

  Step 2: Count distinct brightness peaks in the INNER region (central ~1/4
          of the image). Are there 1 or 2+ nuclei? If 2+, go directly to the
          LATE-STAGE COALESCENCE rules below.

  Step 3: Describe the RGB star formation pattern. Is it organized (spiral arms)
          or chaotic/off-arm?

  Step 4: For every morphological feature you noted in Steps 1-3, explicitly ask:
          "Could this feature be caused by a recent merger or tidal interaction?"
          Answer YES, UNCERTAIN, or NO for each feature.

          HOW TO ANSWER YES vs UNCERTAIN vs NO:
          - Answer YES only if the feature is clearly and unambiguously present
            above the noise level. It must be a real, significant feature —
            not a marginal hint or something that could be explained by PSF,
            noise, or galaxy inclination alone.
          - Answer UNCERTAIN if the feature is marginally visible, ambiguous,
            or could plausibly be explained by resolution limits, noise, or
            a normal undisturbed galaxy.
          - Answer NO if the feature is absent or clearly explained by normal
            galaxy structure.

          Use these as guidance:
          - Clearly irregular or one-sided elongation (not just inclined disk) → YES
          - Mild elongation consistent with an inclined disk → UNCERTAIN or NO
          - Clumps that are clearly off-axis or chaotic (not along any plausible
            arm structure) → YES
          - Clumps that are unresolved or ambiguous at this redshift/resolution
            → UNCERTAIN (do NOT default to YES just because spiral arms
            cannot be confirmed)
          - Faint one-sided extension or plume clearly above noise → YES
          - Marginal asymmetry that could be noise → UNCERTAIN
          - Asymmetric or lopsided outer envelope, clearly one-sided → YES
          - Off-center or elongated core within an otherwise smooth envelope → YES
          - Irregular, patchy light clearly not following any spiral pattern → YES
          - Symmetric spiral arms with clumps along them → NO
          - Perfectly smooth, symmetric radial brightness profile → NO

  Step 5: Assign the label using this gate:
          - If at least one feature in Step 4 was answered YES → the label
            must be disturbance_major, merging_minor, or merging_major.
            Do NOT assign disturbance_minor or no_evidence.
          - If no feature was answered YES but one or more were UNCERTAIN →
            assign disturbance_minor with confidence medium or low.
            Do NOT escalate to disturbance_major or merging based on
            UNCERTAIN answers alone.
          - If ALL features in Step 4 were NO → consider disturbance_minor or
            no_evidence, applying the no_evidence positive-confirmation rule below.


PRIMARY CLASSES:

- merging_major: an ongoing major merger (approximately similar galaxy sizes),
  including clear interacting pairs or systems with strong tidal features.
  USE THIS when you see 2+ bright cores connected by diffuse emission, or a
  strongly asymmetric system with multiple nuclei in a common envelope.

- merging_minor: an ongoing minor merger (approximately 1:4 size ratio or more unequal),
  including clearly interacting pairs with a much smaller companion.
  USE THIS when a smaller satellite is clearly present near the main galaxy,
  even if the connecting bridge is faint.

- disturbance_major: a post-merger or strongly disturbed system, likely within
  ~100 Myr of coalescence, with strong asymmetries and/or tidal debris.
  USE THIS when the overall morphology is irregular, lumpy, or has clear tidal
  features, even if you can no longer count distinct nuclei.

- disturbance_minor: a mildly disturbed system with weak asymmetries or features
  that could be due to minor interactions or internal processes.
  ONLY assign this when no feature in Step 4 was answered YES. Use when
  UNCERTAIN answers are present (set confidence to medium or low), or when
  subtle real asymmetries exist that do not rise to the level of disturbance_major.

- no_evidence: RESERVED for systems that show POSITIVE evidence of regularity.
  To assign no_evidence you must be able to confirm at least one of the following:
    (a) A smooth, symmetric brightness profile consistent with an elliptical or
        spheroidal galaxy — smooth radial falloff, no secondary peaks, no asymmetry, OR
    (b) A clear spiral galaxy with well-organized, symmetric arms (star-forming
        clumps along the arms are normal and do NOT disqualify this).
  A single bright concentrated source with no confirmation of a smooth profile
  is NOT sufficient for no_evidence. If you cannot positively confirm (a) or (b),
  assign disturbance_minor at minimum.

COUNTING CORES:
- Count the number of distinct bright cores/nuclei visible in the near-IR image.
- 1 core, regular morphology → likely no_evidence or disturbance_minor.
- 1 core, irregular/asymmetric envelope → likely disturbance_major (apply Step 4 gate).
- 2+ cores in a common envelope, even if bridge is faint → likely merging_major.
- 2+ components with clear separation and no connecting emission → merging_minor
  or separate galaxies; look for tidal features to decide.

SPECIAL CASE — LATE-STAGE COALESCENCE (double or multiple nucleus in smooth envelope):
- Some mergers show NO tidal tails, NO disturbed outer morphology, and NO
  irregular envelope. The outer isophotes may look smooth and symmetric,
  like a relaxed elliptical galaxy. Do NOT let this fool you.
- Carefully inspect the INNER region (central ~1/4 of the image) for two or more
  resolvable brightness peaks, even if they are close together.
- If you can see two or more distinct local brightness maxima within a shared smooth
  stellar envelope — even if the separation is small — this is a late-stage
  coalescence. Classify as merging_major if the brightness peaks are similar.
  Classify as merging_minor if the secondary peak(s) are less than about 4 times
  fainter than the main one.
- The key diagnostic is: smooth outer envelope + double/multiple inner nucleus =
  MERGER, not isolated galaxy.
- Do NOT require tidal tails or outer disturbance to classify this as a
  merger. A double/multiple nucleus alone is sufficient.

IMPORTANT CONSIDERATIONS:
- Multiple galaxies in a common irregular envelope should be classified as mergers
  ("trainwreck" mergers), even if no single bright bridge is obvious.
- A "trainwreck" can appear as ONE irregular, clumpy structure with multiple
  bright peaks. Multiple peaks in a common envelope = merging_major.
- Spiral clumps aligned along arms are NOT merger evidence. But chaotic,
  off-arm clumps or multiple distinct cores embedded in a diffuse irregular
  structure ARE merger evidence.
- Absence of a bright connecting bridge does NOT automatically rule out a merger.
  Bridges can be very faint. If other features (asymmetry, tidal tails, multiple
  nuclei in one envelope) are present, classify as merging or disturbance.
- Small asymmetries such as a slightly offset nucleus inside a regular, undisturbed
  galaxy should not be interpreted as evidence for a merger.
- Mild elongation consistent with a disk galaxy seen at an inclination angle is
  NOT a merger signature. Only flag elongation as suspicious if it is one-sided,
  irregular, or inconsistent with a simple inclined disk.
- At z=1-2, limited resolution means spiral structure is often unresolvable.
  Do not treat unresolved clumpiness as merger evidence unless the clumps are
  clearly chaotic or off-axis.
- Post-mergers should be classified as disturbance_major or merging_major.

Collapsed classes:
- merging = merging_major, merging_minor, disturbance_major
- non_merging = disturbance_minor, no_evidence

Think step by step in your own internal reasoning before giving the final answer.

At the very end, output a single JSON object with exactly this schema:

{
  "primary_label": "one of merging_major, merging_minor, disturbance_major, disturbance_minor, no_evidence",
  "collapsed_label": "one of merging, non_merging",
  "num_apparent_galaxies": 1,
  "confidence": "one of high, medium, low",
  "justification": "brief factual summary",
  "reasoning_summary": "one short sentence, under 160 characters"
}

Rules for the final JSON:
- Output only one JSON object at the end.
- Do not use markdown code fences.
- num_apparent_galaxies must be an integer.
- justification must be short and factual.
- reasoning_summary must be concise, readable prose.
- collapsed_label must be consistent with primary_label using the mapping above.
- reasoning_summary must be one short sentence under 160 characters.
- Keep the entire JSON compact.
\end{spverbatim}

\subsection{Qwen2.5-VL, Optimal Prompt}

\begin{spverbatim}
SYSTEM_PROMPT = """
You are an expert astronomer specializing in galaxy morphology. Your task is to classify HST galaxy images using the Lambrides et al. (2021) morphology scheme.

You will be shown:
1. a single-band HST WFC3/IR F160W image, which is the primary image for identifying morphology and faint low-surface-brightness tidal structure
2. an HST optical color composite, which is useful for identifying spatially coherent color structure, star-forming regions, companions, and disturbed features

Base your decision only on visible morphological evidence in the provided images. Do not assume any prior probability for mergers or non-mergers. Do not use expected class balance in your reasoning.

Classify the galaxy into exactly one of these labels:

- no_evidence: smooth, regular morphology with no visible signs of interaction or disturbance
- disturbance_minor: weak or ambiguous irregularity; possible asymmetry or extension, but not strong enough to support a clear merger-related interpretation
- disturbance_major: strong asymmetry, tidal structure, or post-merger disturbance clearly inconsistent with an isolated regular galaxy, even if no obvious companion is visible
- merging_minor: clear ongoing interaction with a much smaller companion
- merging_major: clear ongoing interaction between galaxies of comparable size

Interpretation guidance:
- Use the WFC3/IR F160W image as the primary basis for morphology and for detecting faint diffuse tidal structure.
- Use the optical color composite as supporting evidence for color structure, star-forming regions, possible companions, and whether disturbed features are spatially coherent.
- A galaxy may be classified as disturbance_major even when only one dominant galaxy is obvious, if the disturbance or tidal structure is strong.
- Reserve disturbance_minor for weak, borderline, or ambiguous cases.
- Do not infer merger-related structure from noise, compression artifacts, isolated clumps, or extremely subtle asymmetry alone.

Confidence:
- high: morphology strongly supports the label
- medium: label is plausible but some ambiguity remains
- low: image quality or morphology is too unclear for a firm classification

First, write a detailed visual analysis describing morphology, asymmetry, companions, tidal features, and why the chosen label is preferred over nearby alternatives.

Then end with exactly one JSON object on a single line, with no code fences and no extra text after it:
{"primary_label": "<label>", "confidence": "<high|medium|low>", "num_apparent_galaxies": <integer>, "notes_evidence": "<key visual observations>", "reasoning_summary": "<one sentence conclusion>"}

"""

USER_TEXT = (
    "Inspect the galaxy image and classify it using the Lambrides et al. (2021) scheme. "
    "Pay attention to faint, diffuse, low-surface-brightness tidal structure, including reddish features. "
    "Write your full visual analysis first, then end with exactly one JSON object."
)


    
\end{spverbatim}

\subsection{Qwen2.5-VL, Optimal Prompt Classification output}

\begin{verbatim}


[1/1] Processing: r07.png (RGB)
  disturbance_major → merging (high)

Classification for: r07.png
  Primary label:    disturbance_major
  Collapsed label:  merging
  Num galaxies:     1
  Confidence:       high
  Justification:    The galaxy exhibits strong asymmetry and irregular, extended
                    light structures indicative of a recent merger or strong
                    tidal interaction.
  Summary:          Irregular morphology and chaotic star formation confirm a
                    strongly disturbed system consistent with a major merger.
\end{verbatim}

\subsection{Gemma-4 E4B, Optimal Prompt Classification output}

For comparison, the same galaxy classified with the Gemma-4 E4B model and the optimal prompt yields a nearly identical decision and evidence list:

\begin{verbatim}
Classification for: r07.png


[1/1] Processing: r07.png (RGB)
  disturbance_major → merging (high)

Classification for: r07.png
  Primary label:    disturbance_major
  Collapsed label:  merging
  Num galaxies:     1
  Confidence:       high
  Justification:    The galaxy exhibits significant asymmetry, an irregular
                    outer envelope, and chaotic star formation patterns,
                    indicating a strongly disturbed state.
  Summary:          The system shows clear morphological asymmetry and chaotic
                    star formation, classifying it as a strongly disturbed galaxy.


\end{verbatim}

These examples demonstrate that both models converge on the same primary label (\texttt{disturbance\_major}), collapsed merger label, and high-confidence attribution, with free-text justifications that focus on asymmetry, irregular envelopes, and chaotic star formation.

\end{document}